\documentclass[sn-nature,Numbered]{sn-jnl}
\usepackage{xr}
\usepackage{xr-hyper}
\usepackage{lmodern}
\usepackage{hyperref}
\makeatletter
\usepackage{graphicx}%
\usepackage{multirow}%
\usepackage{amsmath,amssymb,amsfonts}%
\usepackage{amsthm}%
\usepackage{mathrsfs}%
\usepackage[title]{appendix}%
\usepackage{xcolor}%
\usepackage{textcomp}%
\usepackage{manyfoot}%
\usepackage{booktabs}%
\usepackage{algorithm}%
\usepackage{algorithmicx}%
\usepackage{algpseudocode}%
\usepackage{listings}%
\usepackage{subfigure}

\begin{document}

\title{Affective Polarization and Dynamics of Information Spread in Online Networks}

\author[1,2,*]{Kristina Lerman}
\author[2]{Dan Feldman}
\author[2]{Zihao He}
\author[2]{Ashwin Rao}
\affil[1]{USC Information Sciences Institute, Marina del Rey, CA, USA}
\affil[2]{USC, Computer Science Dept, Los Angeles, CA,  USA}

\affil[*]{lerman@isi.edu}

\abstract{

Members of different political groups not only disagree about issues but also dislike and distrust each other. While social media can amplify this emotional divide---called affective polarization by political scientists---there is a lack of agreement on its strength and prevalence. 
We measure affective polarization on social media by quantifying the emotions and toxicity of reply interactions. We demonstrate that, as predicted by affective polarization,  interactions between users with same ideology (in-group replies) tend to be positive, while interactions between opposite-ideology users (out-group replies) are characterized by negativity and toxicity. Second, we show that affective polarization generalizes beyond the in-group/out-group dichotomy and can be considered a structural property of social networks. Specifically, we show that emotions vary with network distance between users, with closer interactions eliciting positive emotions and more distant interactions leading to anger, disgust, and toxicity. Finally, we show that similar information exhibits different dynamics when spreading in emotionally polarized groups. These findings are consistent across diverse datasets  spanning discussions on topics such as the COVID-19 pandemic and abortion in the US. Our research provides insights into the complex social dynamics of affective polarization in the digital age and its implications for political discourse.
}

\flushbottom
\maketitle

\thispagestyle{empty}

\section{Introduction}

Democrats and Republicans in the United States not only disagree on many economic, political and cultural issues but have also grown less tolerant of opposing viewpoints, with members of each party disliking and distrusting those affiliated with the opposing party. The emotional divide---dubbed \textit{affective polarization} by political scientists~\cite{iyengar2012affect,iyengar2019origins}---has become a destabilizing force in a democracy, 
reducing cooperation across party lines, stoking hostility towards out-group members~\cite{kingzette2021affective,whitt2021tribalism,rudolph2021affective,berntzen2023consequences} and eroding trust in experts and institutions. For example, the polarized response to the COVID-19 pandemic led individuals affiliated with one political party to distrust recommendations from public health experts if they were supported by the opposing party, hindering effective response to the health crisis~\cite{green2020elusive,grossman2020political,gollwitzer2020partisan}.

Research shows that affective polarization is driven by factors including the news media, political elites, and demographics~\cite{iyengar2019origins,webster2017ideological,whitt2021tribalism}.
Although scholars disagree on how much social media contributes to  partisan animosity~\cite{nordbrandt2022affective}, many see it as an important amplifier~\cite{tornberg2021modeling}. Social media discourse tends to promote inflammatory language and moral outrage directed at the out-group~\cite{brady2017emotion,brady2021social,rathje2021out}. Moreover, social media echo chambers, which segregate users within communities of like-minded others, may amplify political polarization by exposing users to extreme and divisive content~\cite{cinelli2021echo}. Beyond echo-chambers, exposure to out-party views exists~\cite{bakshy2015exposure} and may worsen political polarization~\cite{bail2018exposure}.

\begin{figure}[ht]
    \centering
    \includegraphics[width=0.5\linewidth]{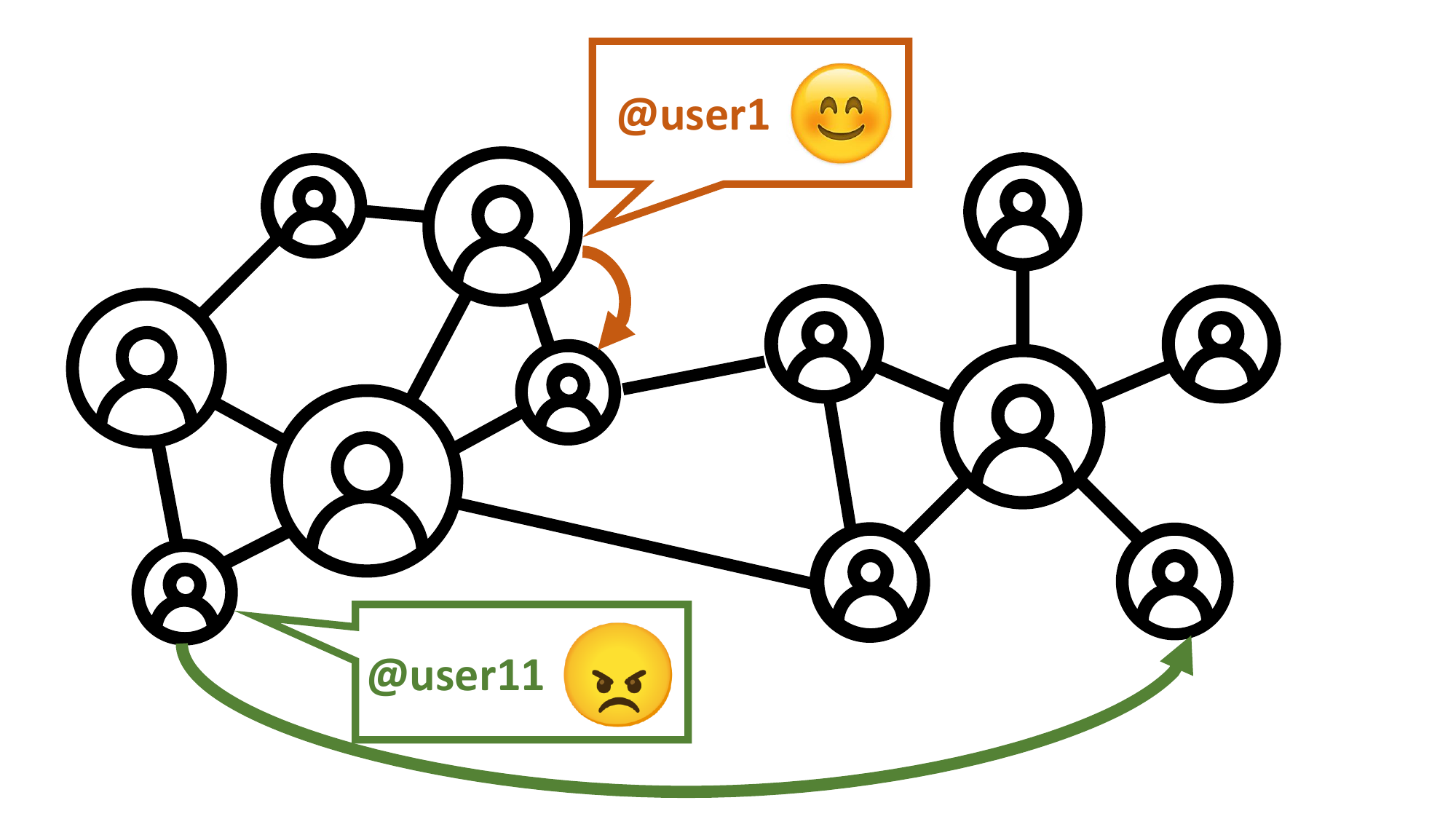}
    \caption{Affective polarization in networks. People express warmer feelings when replying to those closer to them in a social network; when replying to those farther away, they express more negativity.}
    \label{fig:illustration}
\end{figure}

This paper describes an empirical study that makes a three-fold contribution to this body of research: we propose an instrument to quantify affective polarization in social media, establish an empirical relationship between emotions and structure of online networks, and characterize partisan differences in the dynamics of information spread on them.
In contrast to existing research, which measures affective polarization by looking at how people talk \textit{about} out-group members~\cite{tyagi2020affective,yarchi2021political,mentzer2020measuring,rathje2021out}, we focus on how they talk \textit{to} them. We leverage state-of-the-art language models to measure emotions and toxicity of online interactions between social media users and demonstrate they exhibit the hallmarks of affective polarization~\cite{iyengar2012affect}, namely \textit{in-group favoritism}--\textit{out-group animosity}.Moreover, these emotions vary  with distance between interacting users in a social network, demonstrating an association between affective polarization and the structure of online networks. This idea is illustrated  in Figure~\ref{fig:illustration}: when replying to people closer to them in a network (e.g., a retweet network), users express more positive emotions, but when replying to those who are farther away, they express more negative emotions and toxicity. These findings are consistent across datasets of discussions about the COVID-19 pandemic~\cite{chen2020tracking} and abortion~\cite{chang2023roeoverturned} on Twitter. Finally, we analyze the spread of information and show that discussions of contentious issues within partisan groups exhibit different dynamics.
Some issues show random bursts of re-sharing, consistent with being driven by the news cycle, while others persist over longer periods of time, reflecting how  their emotional salience to ideological divisions helps focus attention of polarized groups.

Our study sheds light on the complex dynamics of affective polarization in the digital age, offering insights into the emotional foundations of political discourse on social media and the interaction of emotions with network structure and information diffusion.

\begin{table*}[tbh]
    \centering
    \begin{tabular}{lrrrr}
    \toprule
    \textbf{Dataset} &  \textbf{\#Retweets} &  \textbf{\#Retweeters} &  \textbf{\#Replies} &  \textbf{\#Responders} \\
    \midrule
    Roe\_v\_Wade &7,131,980 &1,005,156 &460,868 &172,988 \\
    COVID-19 &46,419,871 &10,758,690 &4,173,679 &833,875 \\
    
    \bottomrule
    \end{tabular}
    \caption{Number of retweets and replies from the datasets in our study. Number of retweeters and the number of responders gives the total number of unique user ids that participated in a retweet and reply interaction respectively.}
    \label{tab:tweet_stats}
\end{table*}

\section{Results}
We study two massive datasets of online discussions on Twitter: tweets about the \textit{COVID-19} pandemic in the US and tweets about the overturning of the \textit{Roe v Wade} decision that legalized abortion in US in 1973. We classify users' ideology as liberal or conservative based on the text of their tweets, and use transformer-based language models to detect emotions and toxic language in their replies (see Methods). Finally, we identify contentious issues in online discussions and study dynamics of re-sharing by each ideological group. Table \ref{tab:tweet_stats} summarizes statistics of the two datasets.

\begin{figure*}[th]
    \centering
     {\includegraphics[width=1.0\linewidth]{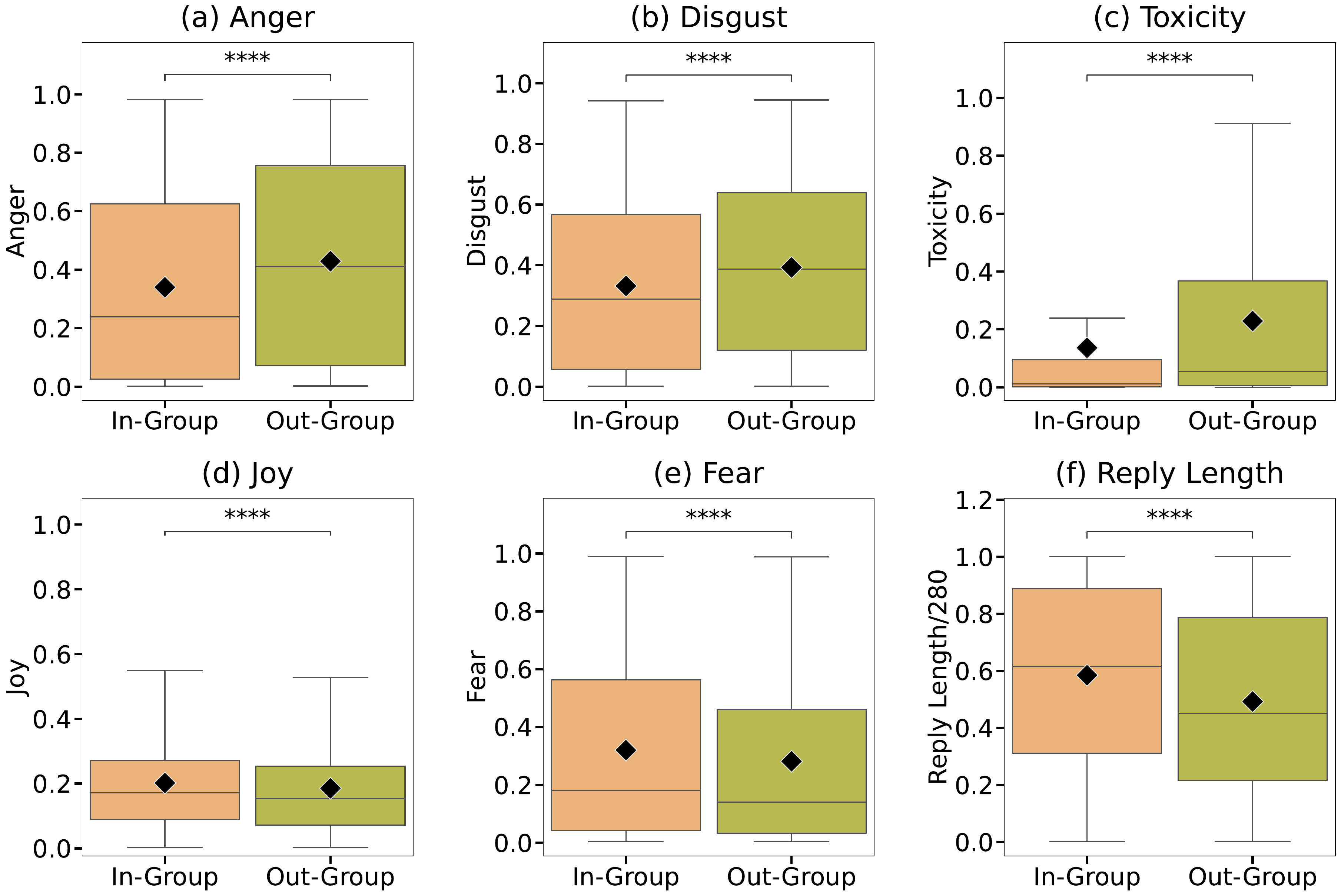}}
    \caption{In-group and out-group affect in the Roe\_v\_Wade data. Boxplots show confidence scores of emotions expressed in replies interactions in the abortion discussions between same-ideology users (in-group) and opposite-ideology users (out-group). Out-group interactions show more (a) anger, (b) disgust, and use more (c) toxic language, but also slightly less (d) joy and (e) fear, and  are generally (f) shorter in length (as measured by the number of characters divided by the maximum allowed length, 280 characters). The boxes span the first to third quartiles, with whiskers extending 1.5 times the interquartile range. The horizontal line inside the box represents the median, and diamond symbol marks the mean. Differences in means were tested for statistical significance using a two-sided Mann-Whitney U Test with the Bonferroni correction: * indicates significance at \textit{p}$<0.05$, ** - \textit{p}$<0.01$, *** - \textit{p}$<0.001$, **** - \textit{p}$<0.0001$ and, ns - not-significant.}
    \label{fig:in-out-rvw}
\end{figure*}

\subsection{Emotions of In-group vs Out-group Interactions}
We measure affective polarization by quantifying emotions expressed in reply interactions between users with a known ideology. \textit{In-group interactions} are replies between same-ideology users: liberal replying to a liberal, etc. \textit{Out-group interactions} are replies between opposite-ideology users: liberal replying to a conservative or \textit{vice versa}. Figure~\ref{fig:in-out-rvw} shows the distribution of emotion confidence scores of replies in the abortion  dataset. 

\begin{figure*}[!htb]
    \centering
     {\includegraphics[width=0.9\linewidth]{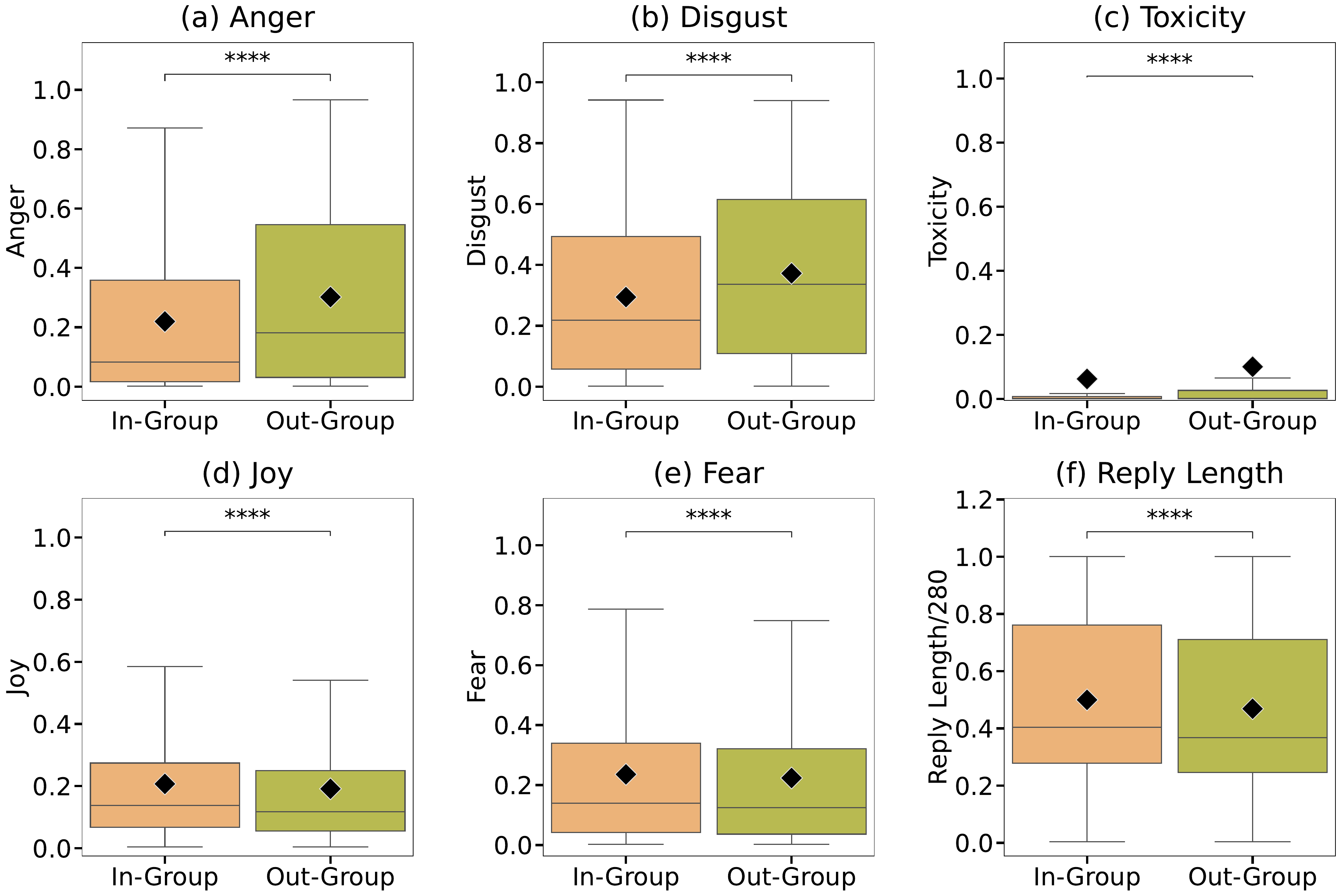}}
    \caption{Affective polarization in COVID-19 discussions: in-group vs out-group affect. Boxplots show confidence scores of emotions expressed in replies between two users with the same ideology (in-group) and users with different ideology (out-group). Out-group interactions show more (a) anger, (b) disgust, and use more (c) toxic language, but also less (d) joy and (e) fear, and tend to be shorter (f). The boxes span the first to third quartiles, with whiskers extending 1.5 times the interquartile range. The horizontal line inside the box represents the median, and diamond symbol marks the mean. Differences in means were tested for statistical significance using a two-sided Mann-Whitney U Test with the Bonferroni correction: * indicates significance at \textit{p}$<0.05$, ** - \textit{p}$<0.01$, *** - \textit{p}$<0.001$, **** - \textit{p}$<0.0001$ and, ns - not-significant.}
    \label{fig:in-out-covid}
\end{figure*}

\begin{figure*}[!hbt]
    \centering
     {\includegraphics[width=1\linewidth]{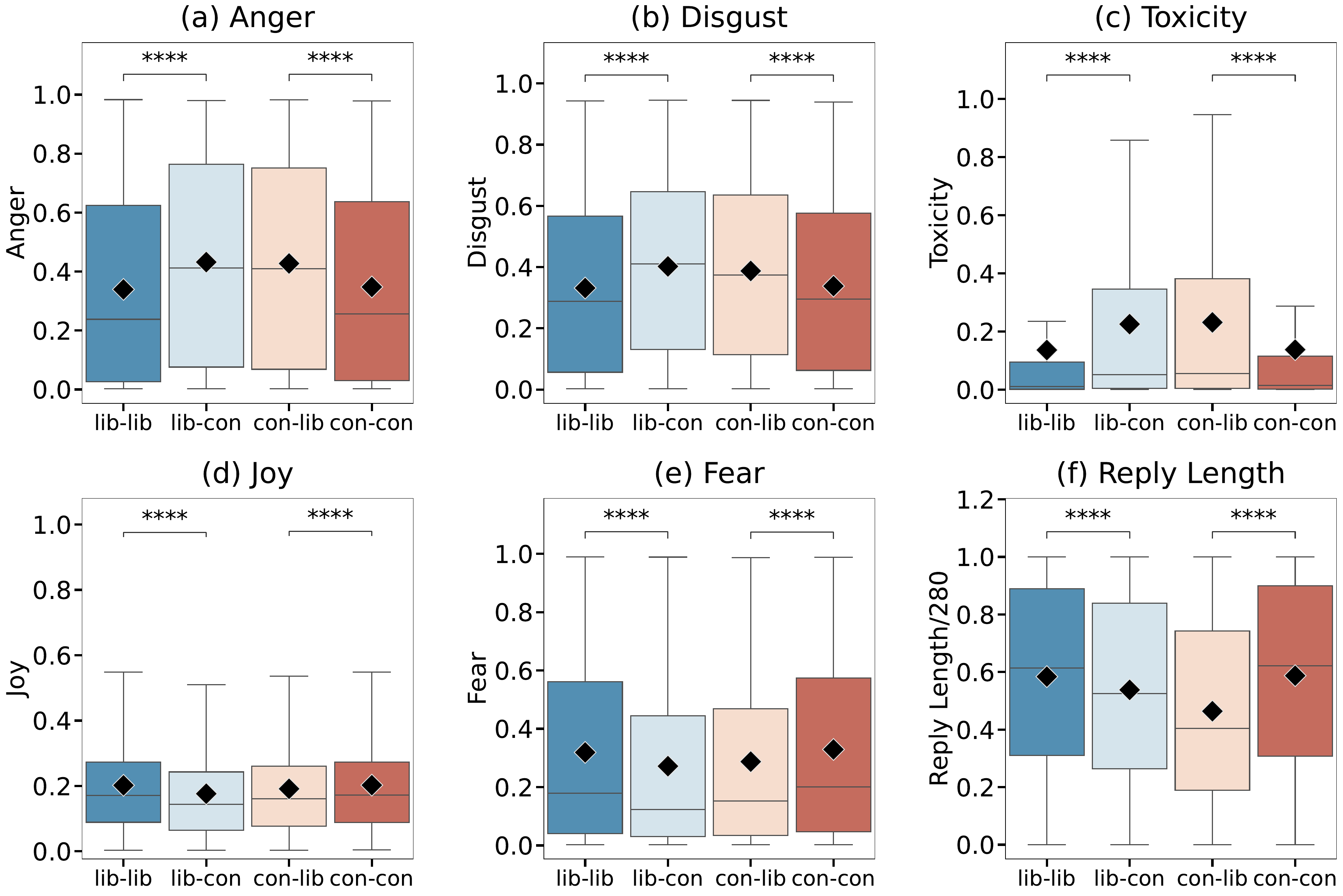}}
    \caption{Partisan differences in affective polarization in the Roe\_v\_Wade abortion discussions. Boxplots show confidence scores for emotions expressed in replies between same-partisanship users (in-group interactions) and opposite-partisanship users (out-group interactions) in the abortion debate, disaggregated by partisanship. Out-group interactions show more (a) anger, (b) disgust, and use more (c) toxic language, but also slightly less (d) joy and (e) fear. Out-group interactions are more likely to be shorter (f), with systematic partisan differences. The boxes span the first to third quartiles, with whiskers extending 1.5 times the interquartile range. The horizontal line inside the box represents the median, and diamond symbol marks the mean. Differences in means were tested for statistical significance using the Mann-Whitney U Test with the Bonferroni correction: * indicates significance at $p<0.05$, ** - $p<0.01$, *** - $p<0.001$, **** - $p<0.0001$ and, $ns$ - not-significant.}
    \label{fig:in-out-lib-con-rvw}
\end{figure*}

\begin{figure*}[!htb]
    \centering
     {\includegraphics[width=1\linewidth]{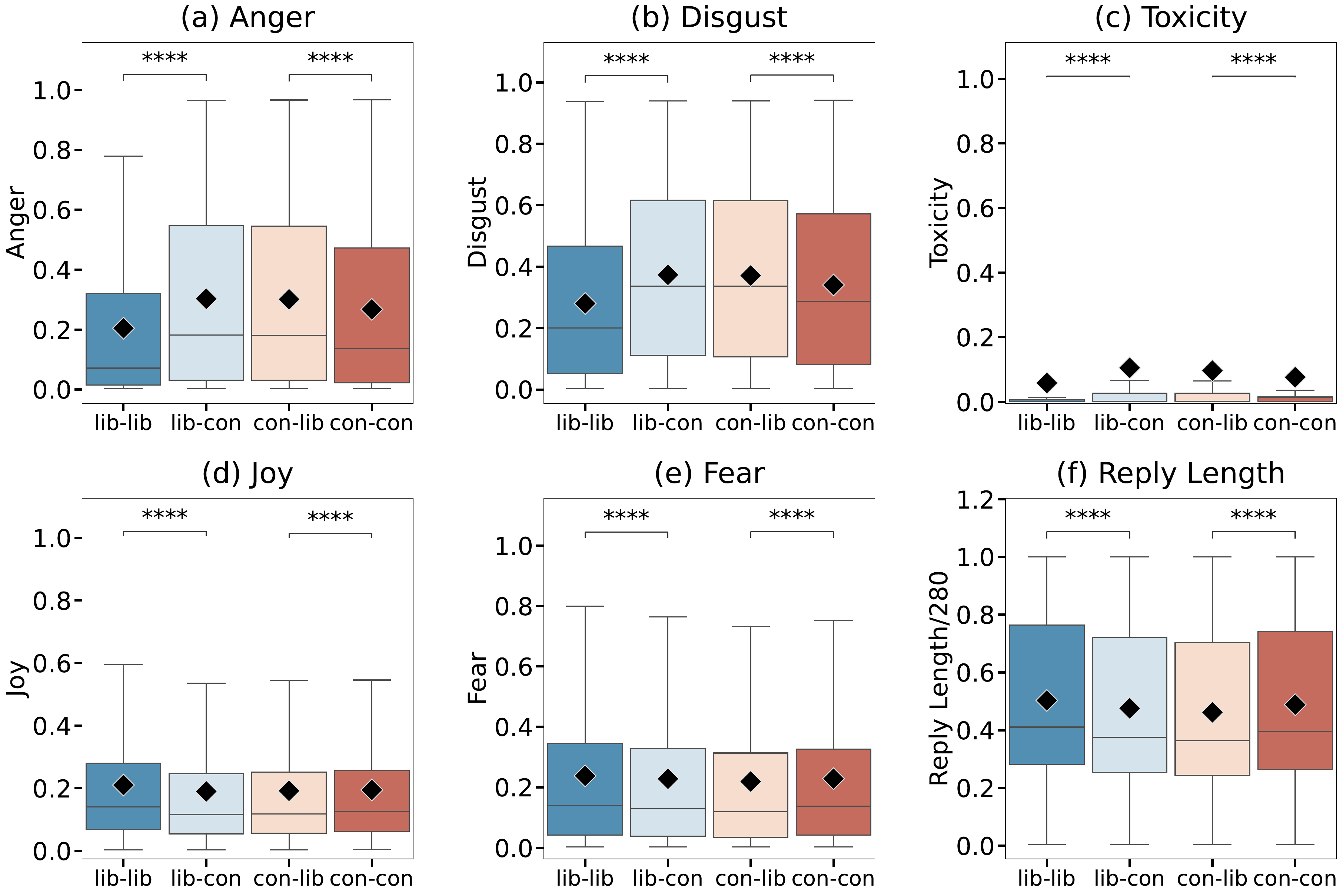}}
    \caption{Partisan differences affective polarization in COVID-19 pandemic discussions. Boxplots show confidence scores for emotions expressed in replies between same-partisanship users (in-group interactions) and opposite-partisanship users (out-group interactions) in the abortion debate, disaggregated by partisanship. Out-group interactions show more (a) anger, (b) disgust, and use more (c) toxic language, but also slightly less (d) joy and (e) fear. Out-group interactions are more likely to be shorter (f), with systematic partisan differences. The boxes span the first to third quartiles, with whiskers extending 1.5 times the interquartile range. The horizontal line inside the box represents the median, and diamond symbol marks the mean.Differences in means were tested for statistical significance using the Mann-Whitney U Test with the Bonferroni correction: * indicates significance at $p<0.05$, ** - $p<0.01$, *** - $p<0.001$, **** - $p<0.0001$ and, $ns$ - not-significant.}
    \label{fig:in-out-lib-con-covid}
\end{figure*}

Anger, disgust and toxicity scores of out-group replies are substantially higher than for in-group replies, consistent with  \textit{out-group animosity}. Similarly, in-group replies have higher scores for joy, consistent with \textit{in-group favoritism}. We observe similar trends in the COVID-19 data (Supplementary Figure~\ref{fig:in-out-covid}).
All differences are statistically significant.
The differences between in-group and out-group interactions remain after disaggregating replies by ideology across both datasets (Supplementary Figure~\ref{fig:in-out-lib-con-rvw} and Supplementary Figure~\ref{fig:in-out-lib-con-covid}).

Our empirical analysis yields additional insights. We did not find significant differences in love, sadness or optimism. Out-group interactions are more emotional but shorter. Additionally, fear does not behave like other negative emotions in that it is higher for in-group interactions. This may reflect the role of fear as an agent of social cohesion. By identifying a common threat (see folktales and legends~\cite{tangherlini2017toward,shahsavari2020conspiracy}), fear creates solidarity, which unites the group.

\begin{figure}[ht]
    \centering
    \includegraphics[width=0.75\linewidth]{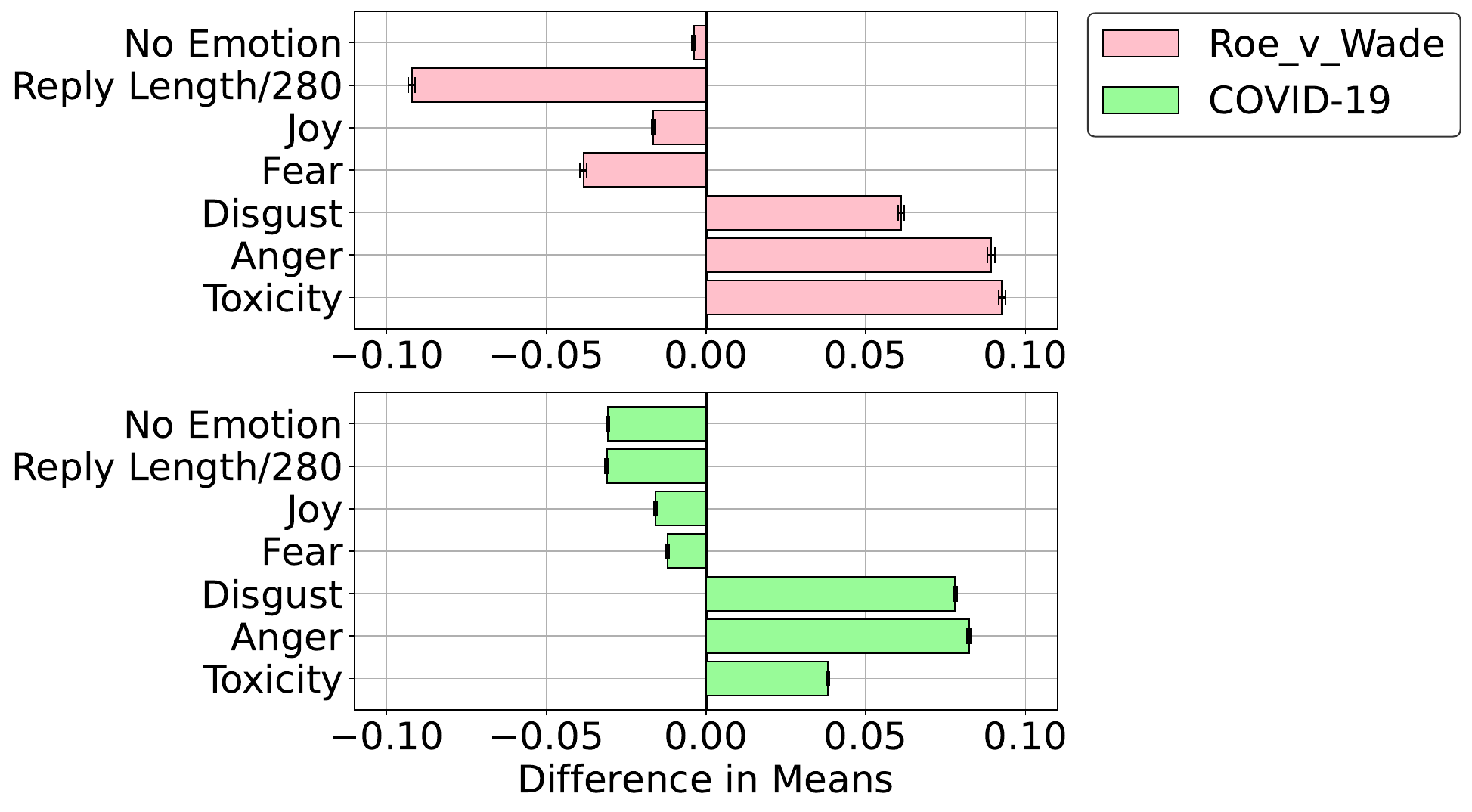}
    \caption{Affective polarization. Difference between the mean emotion confidence scores of out-group and in-group interactions in the Roe\_v\_Wade and COVID-19 datasets. Error bars show standard errors.}
    \label{fig:in-out-diffs}
\end{figure}

Figure~\ref{fig:in-out-diffs} summarizes these results for both datasets by showing the difference between the mean emotion confidence of out-group and in-group replies. There exists a consistent emotion gap: interactions across ideological lines exhibit more negativity and toxicity, and less joy.
Surprisingly, out-group replies are significantly shorter: in the Roe\_v\_Wade data, out-group replies are on average 17 characters shorter, while in the COVID-19 data, they are 8 characters shorter. Together with the finding that out-group interactions are more emotional, this suggests that people communicate across party lines not to convey information but rather to express animosity and to troll.

\begin{figure*}[tph]
    \centering
    \includegraphics[width=1.0\linewidth]{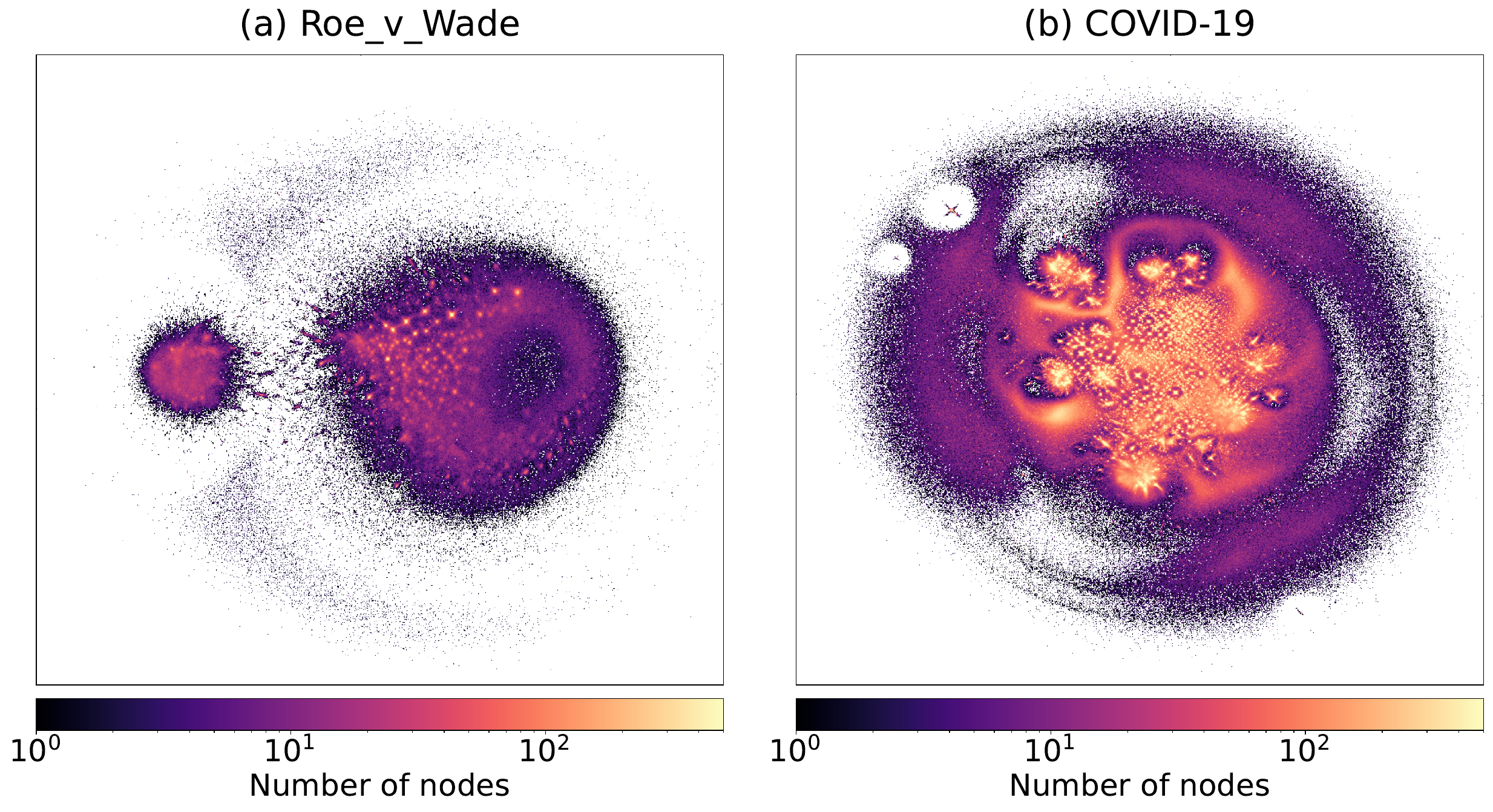}
    \caption{Heatmap of the embeddings of social networks. Each network was constructed by linking users who retweeted each other in online discussions about a) the 2022 overturning of Roe\_v\_Wade and   b) the COVID-19 pandemic. The massive retweet networks were embedded in a lower-dimensional space using a  graph embedding method. The heatmap of the embedding shows bright spots of densely-linked communities of users who frequently retweet one another. The retweet network of abortion discussions (a) shows two overaching polarized communities, while the network of the pandemic discussions (b) has a multi-focal structure. }
    \label{fig:networks}
\end{figure*}

\subsection{Emotions and Network Distance}
\begin{figure*}
    \centering
    {\includegraphics[width=1.0\linewidth]{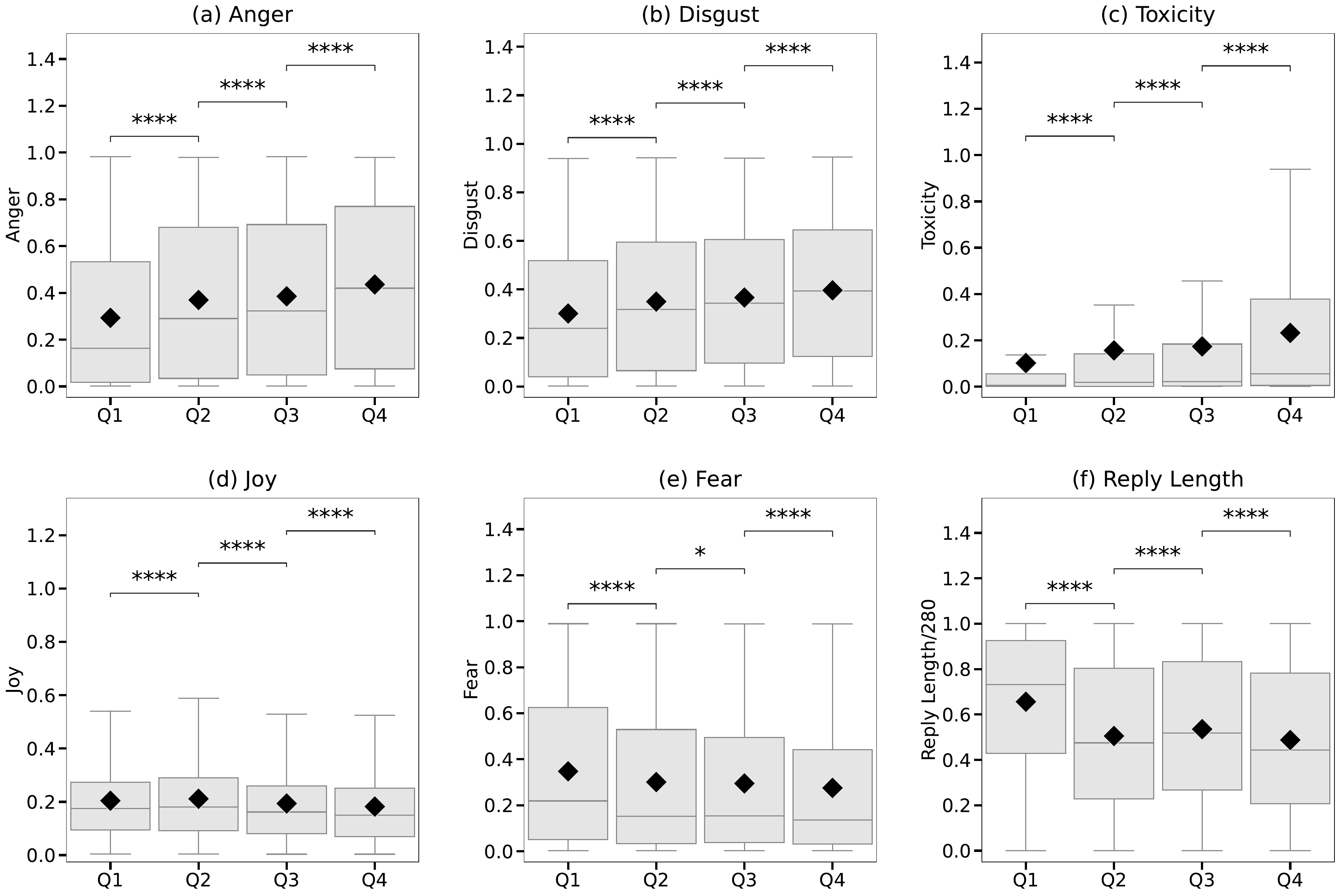}}
    \caption{Affective polarization in the retweet network of Roe\_v\_Wade data. Each plot shows emotions in reply interactions as a function of network distance between interacting users. Network distance is calculated in the network embedding space and then divided into quartiles, with Q1 showing the 25\% of closest interactions and Q4 showing the 25\% of the most distant interactions.  Emotions (a) anger, (b) disgust and  (c) toxicity increase with distance between users in the network embedding space, while (d) joy and (e) fear decrease with distance, as does (f) reply length. The boxes span the first to third quartiles, with whiskers extending 1.5 times the interquartile range. The horizontal line inside the box represents the median, and diamond symbol marks the mean. Differences in means were tested for statistical significance using a two-sided Mann-Whitney U Test with the Bonferroni correction: * indicates significance at \textit{p}$<0.05$, ** - \textit{p}$<0.01$, *** - \textit{p}$<0.001$, **** - \textit{p}$<0.0001$ and, ns - not-significant.}
    \label{fig:distance-lvd-rvw}
\end{figure*}

Next, we explore the interplay between emotions and network structure, using retweet networks to represent the online social networks (see Methods). Due to their large size, we use an embedding technique \textit{LargeVis}~\cite{tang2016visualizing}
to visualize the networks. 
Figure~\ref{fig:networks} shows the heatmap of these embeddings, with bright spots corresponding to dense clusters of users who frequently retweet one another. 

The retweet network of the discussions in the Roe\_v\_Wade data  (Figure~\ref{fig:networks}a) shows two main clusters with additional substructure near the larger cluster. The coarse-grained structure reflects the polarized nature of the abortion debate: most of the liberals are in the larger  cluster and the conservatives are in the smaller one. The retweet network of the COVID-19 discussions (Figure~\ref{fig:networks}b) contains many small clusters. Although these discussions did grow to be polarized, during the first months of the pandemic covered by our dataset, these divisions do not appear entrenched. 

We measure network distance between two users in the retweet network embedding space (see Methods). Figure~\ref{fig:distance-lvd-rvw} shows confidence scores of emotions in reply interactions as a function of network distance between interacting users in the Roe\_v\_Wade data. We divide the distances into four equal-statistics bins, or quartiles. The first quartile (Q1) represents the 25\% of the closest pairs of interacting users, while Q4 represents the 25\% of the most distant pairs. There are systematic differences in emotions: when users reply to those farther away in the retweet network, they tend to express more anger and disgust, and use more toxic language; while they tend to express more joy and fear in replies to closer users and also use more words. All differences between the quartiles are statistically significant. 

\begin{figure}[tbh]
    \centering
    \includegraphics[width=0.75\linewidth]{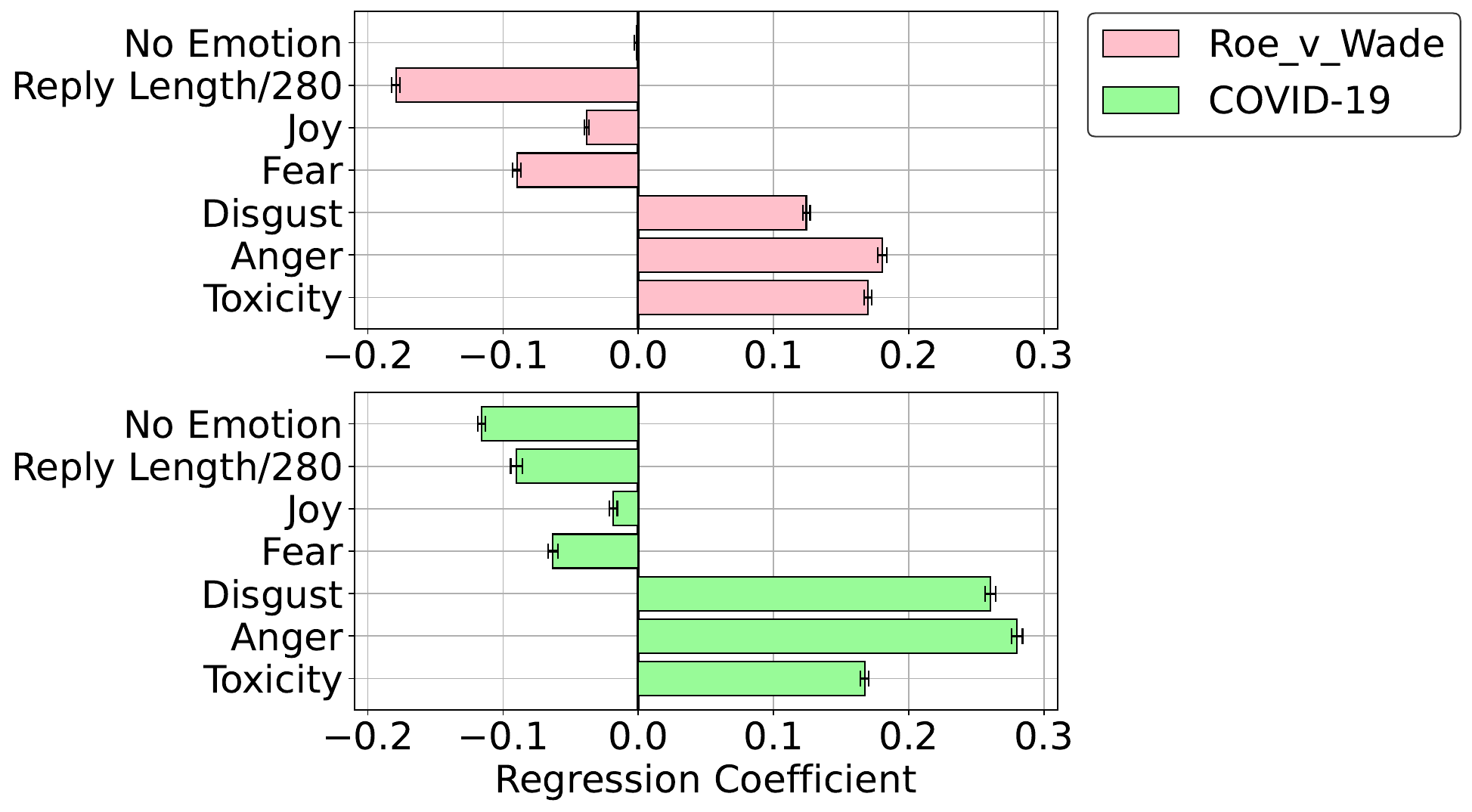}
    \caption{Regression coefficients of affect as a function of network distance in the Roe\_v\_Wade and COVID-19 datasets. The bars represent the value of the regression coefficient of the emotion or toxicity of replies as a function of normalized distance between interacting users in the network embedding space. Distance was normalized by rescaling distances in the embedding space to unit interval. Error bars show 95\% confidence interval. 
    }
    \label{fig:slopes}
\end{figure}

To measure these trends, we perform linear regression on emotion confidence scores using network embedding distance as the independent variable (see Supplementary Figures \ref{fig:distance-rvw} and \ref{fig:distance-covid}). 
Figure~\ref{fig:slopes} brings together all regression coefficients across both datasets, showing similar trends across datasets, in agreement with Figure~\ref{fig:in-out-diffs}. Refer to Supplementary Table \ref{tab:covid_us_full} and Supplementary Table~\ref{tab:roe_wade_us_full} for detailed regressions results. 

\begin{figure*}
    \centering
     {\includegraphics[width=0.95\linewidth]{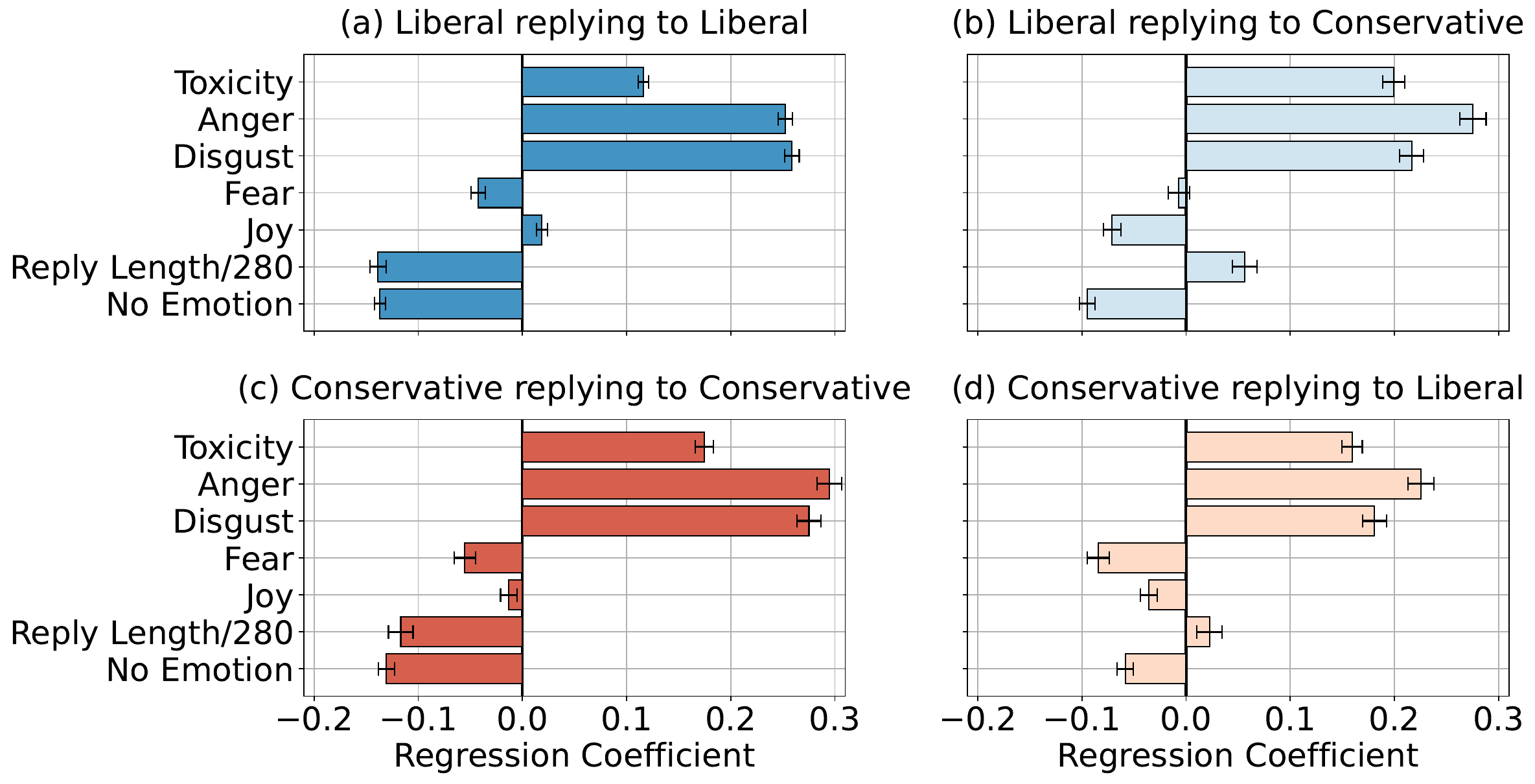}}
    \caption{Regression coefficients of affect as a function of network distance in the retweet network of COVID-19 data. The bars represent the value of the regression coefficient of the emotion or toxicity of replies as a function of embedded RT network distance between interacting users of specific ideologies. For regression analysis, distances were rescaled to unit interval. Error bars show 95\% confidence interval.}
    \label{fig:lvd-slopes-covid}
\end{figure*}

\begin{figure*}
    \centering
     {\includegraphics[width=0.95\linewidth]{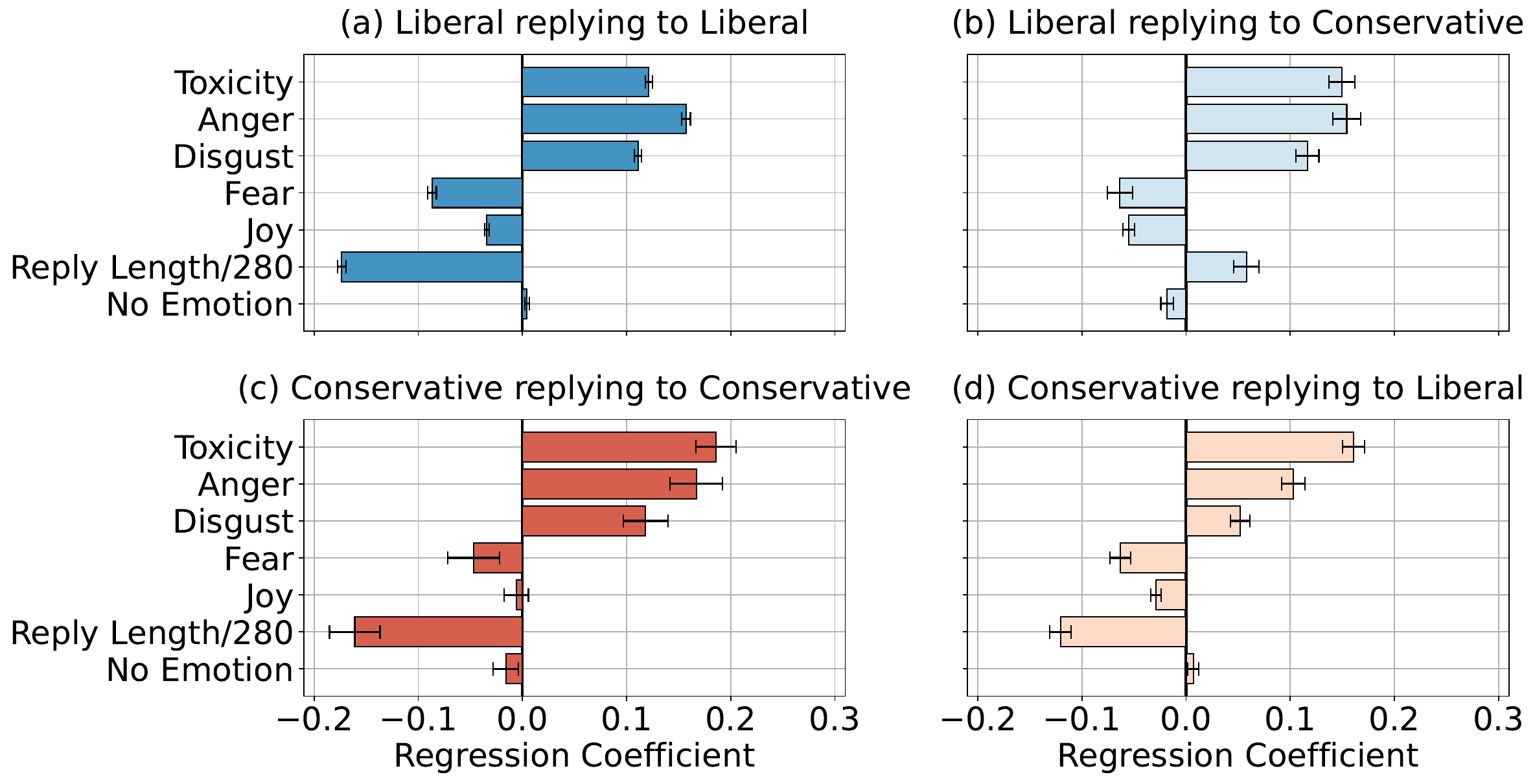}}
    \caption{Regression coefficients of affect as a function of network distance in the retweet network of Roe\_v\_Wade data. The bars represent the value of the regression coefficient of the emotion or toxicity of replies as a function of embedded RT network distance between interacting users of specific ideologies. For regression analysis, distances were rescaled to unit interval. Error bars show 95\% confidence interval.}
    \label{fig:lvd-slopes-rvw}
\end{figure*}

The relationship between emotional expression and network distance holds separately for in-group and out-group interactions and after disaggregating by ideology (Supplementary Figures~\ref{fig:lvd-slopes-rvw}, \ref{fig:lvd-slopes-covid}). 
These results show that we can measure affective polarization even in the absence of partisan labels that define the out-group. Emotions and networks interact so that users feel warmer towards those who are closer to them and colder towards those who are farther away, regardless of group affiliation.

\begin{figure*}
    \centering
     {\includegraphics[width=0.9\linewidth]{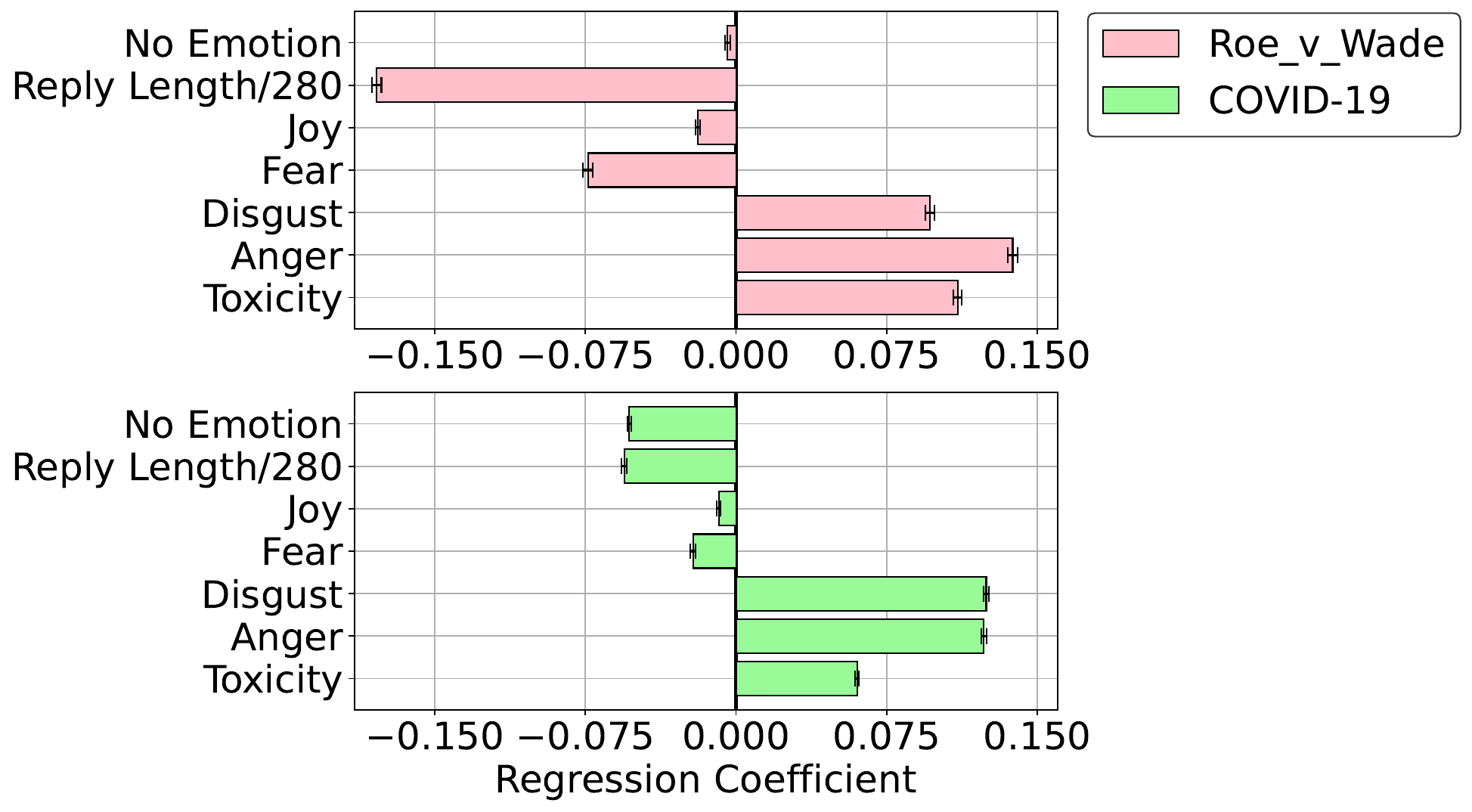}}
    \caption{Regression coefficients of affect as a function of network distance. The bars represent the value of the regression coefficient of the emotion or toxicity of replies as a function of shortest path between interacting users. For regression analysis, path lengths were rescaled to unit interval. Error bars show 95\% confidence interval.}
    \label{fig:spd-slopes}
\end{figure*}

\begin{figure*}
    \centering
     {\includegraphics[width=0.95\linewidth]{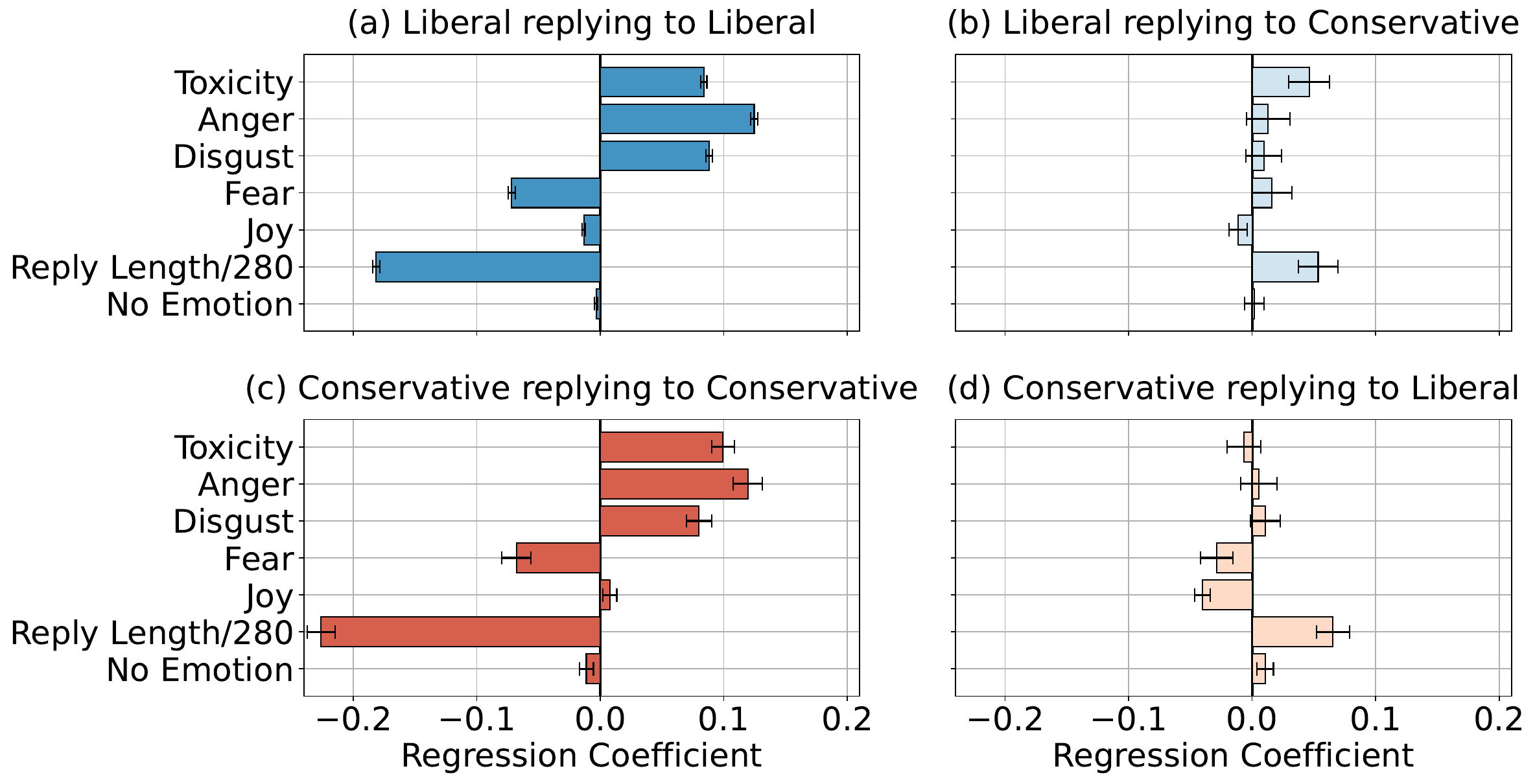}}
    \caption{Regression coefficients of affect as a function of network distance in the retweet network of Roe\_v\_Wade data. The bars represent the value of the regression coefficient of the emotion or toxicity of replies as a function of shortest path between interacting users of specific ideologies. For regression analysis, path lengths were rescaled to unit interval. Error bars show 95\% confidence interval.}
    \label{fig:spd-slopes-rvw}
\end{figure*}

\begin{figure*}
    \centering
     {\includegraphics[width=0.95\linewidth]{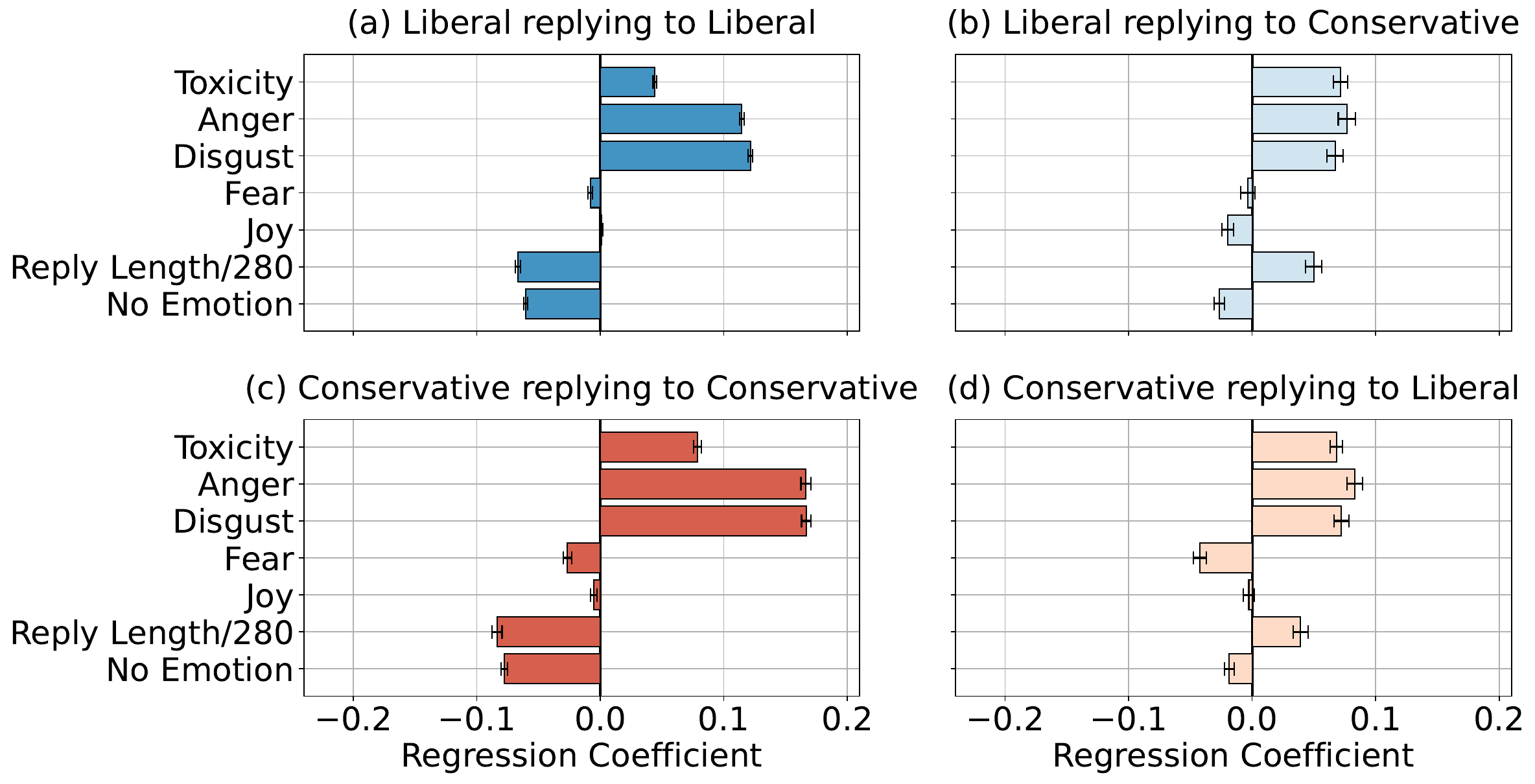}}
    \caption{Regression coefficients of affect as a function of network distance in the retweet network of COVID-19 data. The bars represent the value of the regression coefficient of the emotion or toxicity of replies as a function of shortest path between interacting users of specific ideologies. For regression analysis, path lengths were rescaled to unit interval. Error bars show 95\% confidence interval.}
    \label{fig:spd-slopes-covid}
\end{figure*}

For robustness, we calculate the shortest path length between pairs of nodes in the retweet network as  an alternate measure of network distance. We find that the trends do not depend on how we measure network distance, with results largely remaining the same when using either metric (Supplementary Figure~\ref{fig:spd-slopes}), even after disaggregating by ideology (Supplementary Figures~\ref{fig:spd-slopes-rvw} and  \ref{fig:spd-slopes-covid}).

\subsection{Information Spread in Affectively Polarized Populations}

\begin{figure*}
    \centering
    \begin{tabular}{cc}
    \includegraphics[width=0.46\linewidth]{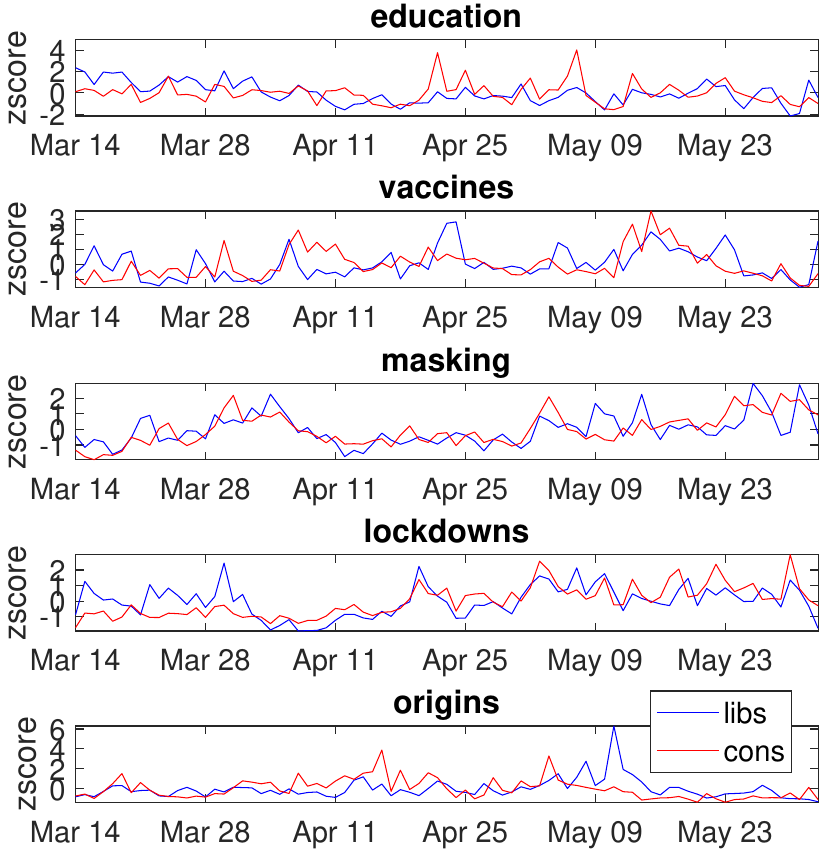} &
    \includegraphics[width=0.46\linewidth]{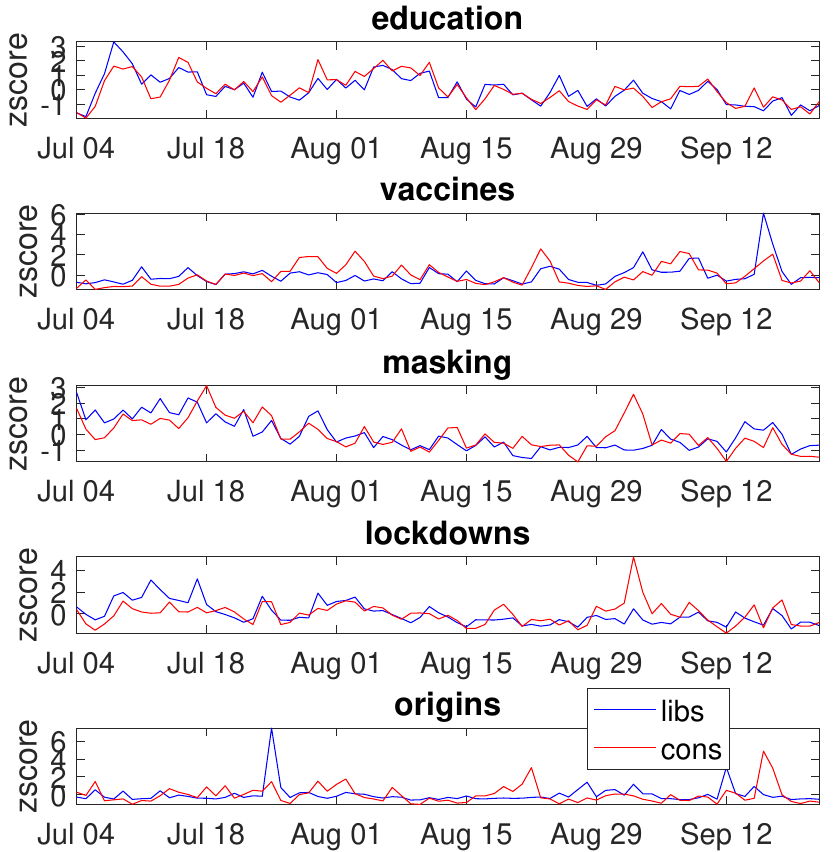} 

    \\
    \includegraphics[width=0.46\linewidth]{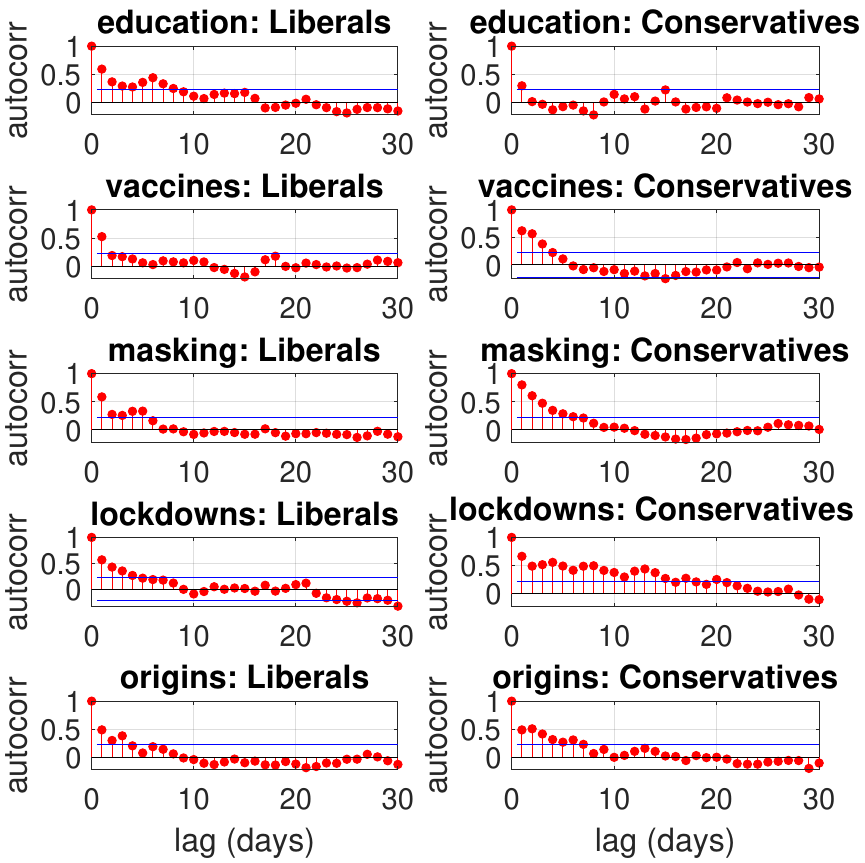} &
    \includegraphics[width=0.46\linewidth]{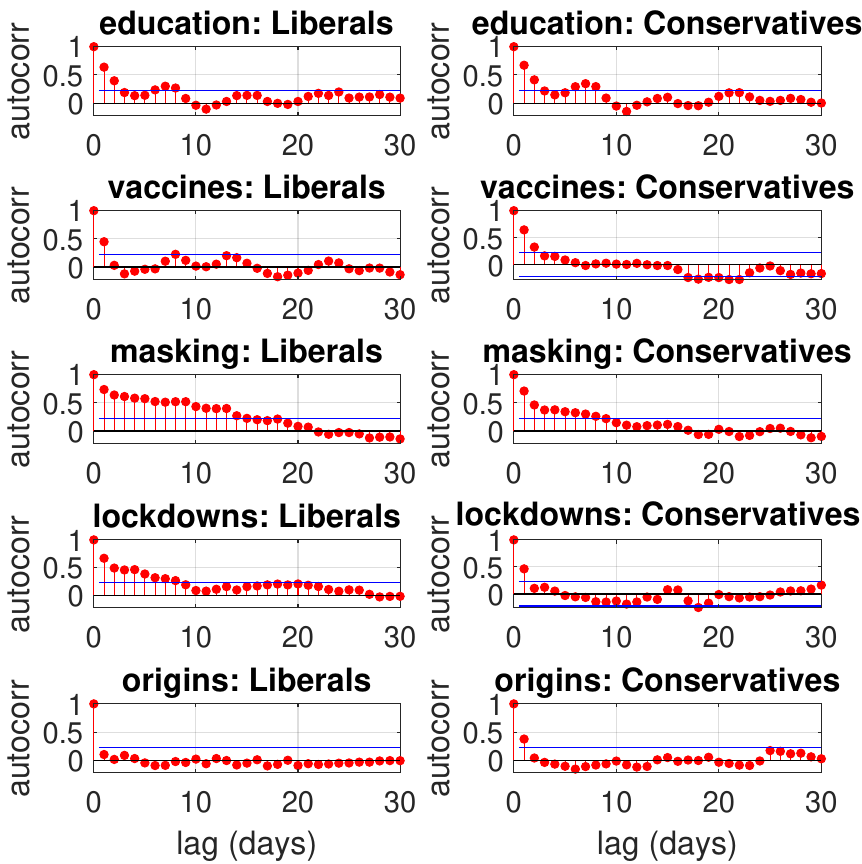} 
    \\
    \includegraphics[width=0.46\linewidth]{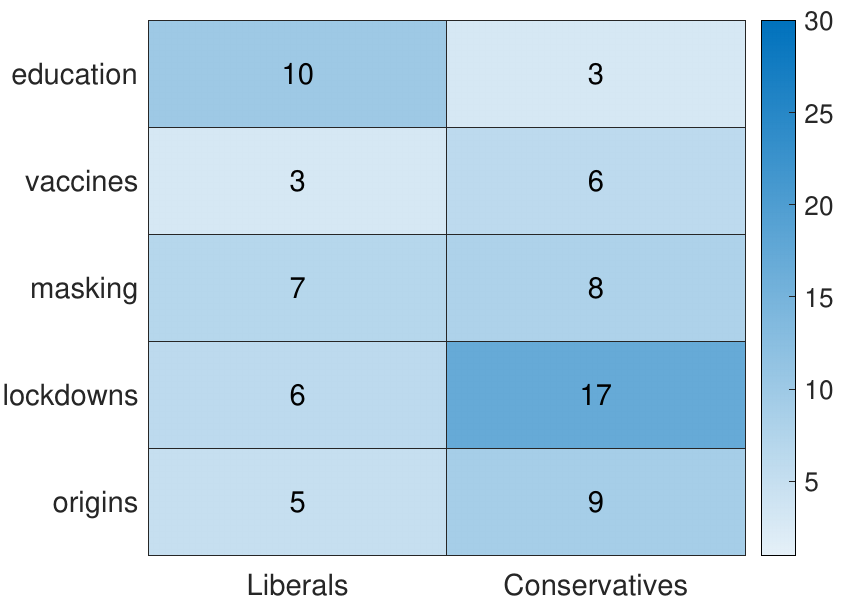} &
    \includegraphics[width=0.46\linewidth]{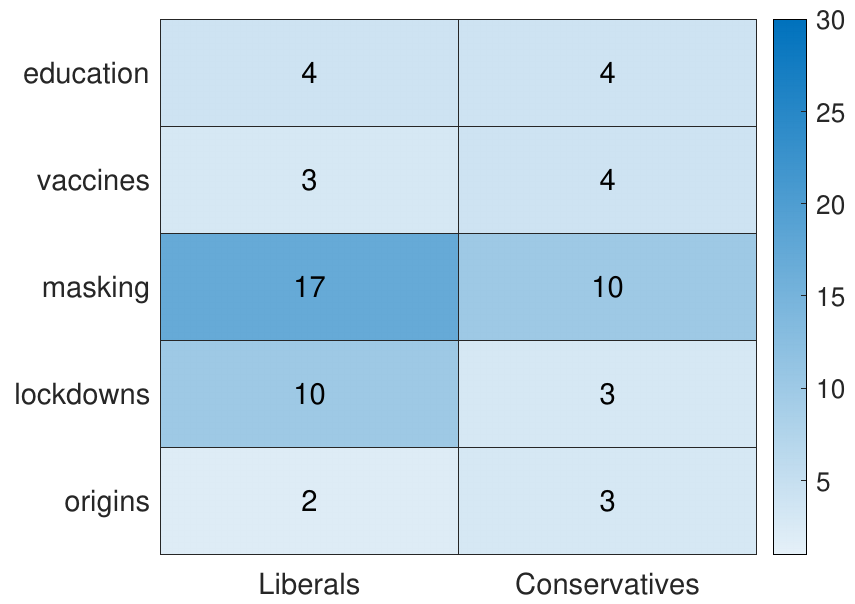} 
\\
Post National Emergency & Summer 2020
    \end{tabular}
    \caption{Persistent dynamics of information spread in COVID-19 discussions. The figures show (top) the time series of the volume of retweets on each issue made by liberals and conservatives, (middle) the autocorrelation function of each time series, and (bottom) the longest significant time lag, in days, of the autocorrelation function. Each column represents a different 80 day time period, (left)  after President Trump's declaration of national emergency, and (right) July 4th holiday. The rows represent pandemic-related issues: education and online learning, vaccines, masking, lockdowns and social distancing, and origins of the coronavirus. }
    \label{fig:covid-dynamics}
\end{figure*}

\begin{figure*}[tbph]
    \centering
    \begin{tabular}{cc}
        \includegraphics[width=0.46\linewidth]{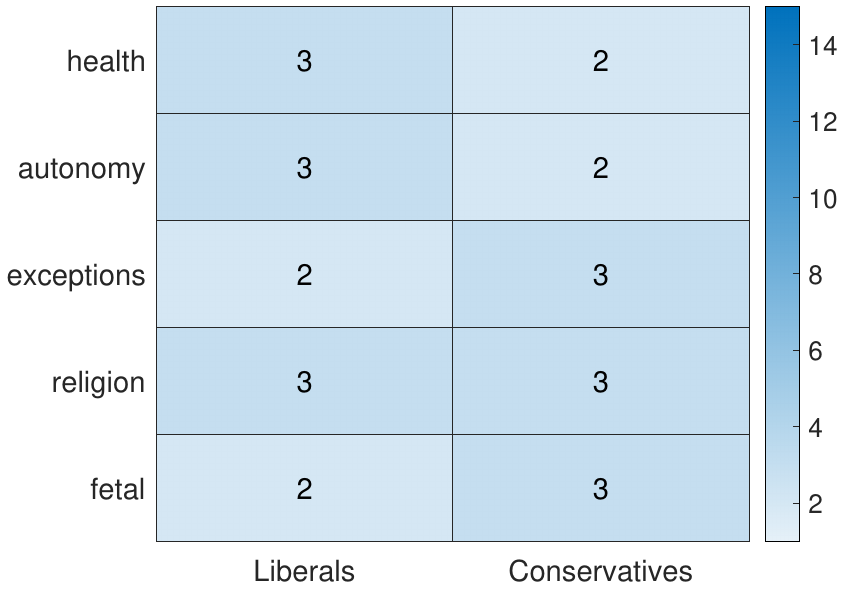} &
        \includegraphics[width=0.46\linewidth]{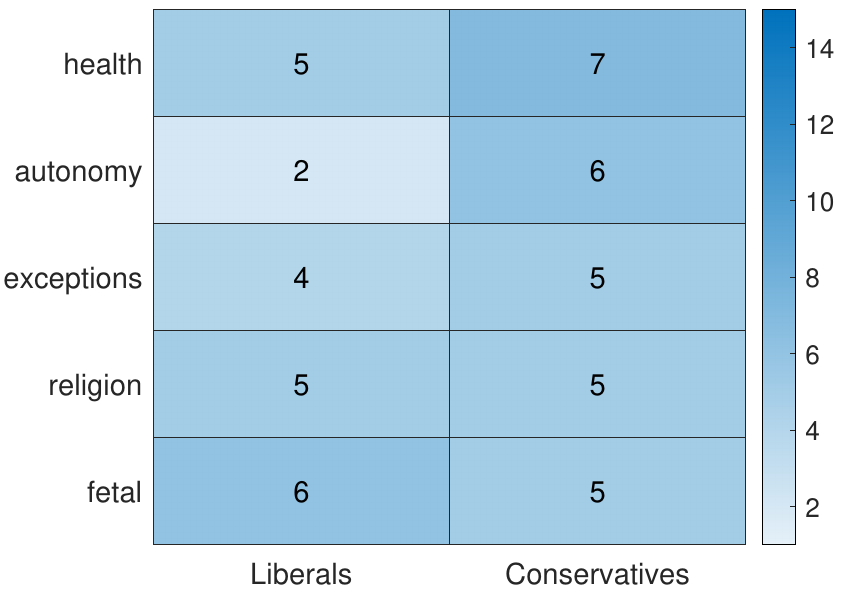}
        \\        Before the leak &  Post  Dobbs ruling 

    \end{tabular}
    \caption{Dynamics of information spread in abortion discussions. The figures show the longest significant time lag, in days, of the autocorrelation function. Each column represents a different 80 day time period, (left)  before the leak of the Dobbs decision, (right) after after SCOTUS's Dobbs decision. The rows represent abortion issues, such as women's health, bodily autonomy, exceptions to abortion bans, religion, and fetal rights.}
    \label{fig:RoevWade_rt_80}
\end{figure*}

Users interact across diverse network distances, seeing information shared by allies and foes alike. Their emotional reactions modulate attention to contentious issues, affecting how those issues spread through the network. Roughly half of the tweets in the COVID-19 corpus discuss at least one contentious \textit{issue}, such as vaccines, education, masking, lockdowns, or origins of the virus (see Methods).  Attention to issues waxes and vanes as events 
drive partisan interest and the diffusion of tweets. 
Figure~\ref{fig:covid-dynamics} (top row) plots the daily number of retweets of each issue by liberals and conservatives at different periods of the pandemic.  The absolute number of retweets varies greatly due to difference in the size of the partisan groups. To address the imbalance we standardize the number of retweets within each group using z-score normalization.

Each time series in Figure~\ref{fig:covid-dynamics} represents the complex dynamics of information diffusion within a group. To  characterize these dynamics, we calculate the autocorrelation function (ACF), which measures the correlation of different points in the same time series, separated by time lags. The middle row of Figure~\ref{fig:covid-dynamics} shows ACF  along with confidence intervals (blue lines).Some ACF plots show weekly patterns in the volume of retweets with peaks at 7, 14, etc. days. However, others show a rapid or gradual decay of ACF to non-significant values. The former trend is consistent with short-lived spikes in retweets occurring at random times, while the latter trend is indicative of persistent attention.
The bottom row of Figure~\ref{fig:covid-dynamics} shows the time lags at which the ACF drops below the confidence interval. 

The early pandemic (post-President Trump's declaration of national emergency, left column of Figure~\ref{fig:covid-dynamics}) was characterized by school closures and challenges of online learning, a topic favored by liberals. Ending lockdowns and reopening the economy became an important issue for conservatives, culminating in protests at state capitals in April 2020. Consistently, discussions about education were persistent among liberals, while  discussions about lockdowns were more persistent among conservatives.  

The summer of 2020 (right column of Figure~\ref{fig:covid-dynamics}) saw mass protests for racial justice, sparked by the murder of George Floyd, in which many liberals participated. Since demonstrations required violating social distancing measures enacted to limit the spread of the virus (lockdowns issue), liberals promoted masking to stay safe, which made masking an important issue among both liberals and conservatives.

In June 2022, the Supreme Court reversed federal guarantees on abortion access made by the 1973 Roe v Wade decision. The decision, as well as the leak of its draft in May 2022, sparked debates on all abortion issues: women's health, bodily autonomy, exceptions to abortion bans, religion, and fetal rights.. The difference in the nature of information spread is evident in Figure~\ref{fig:RoevWade_rt_80}, which compares the autocorrelation function of the time series of the daily volume of retweets in the 80-day period before the leak and after the decision. Before the leak (left column of Figure~\ref{fig:RoevWade_rt_80}), the irregular bursts of retweets, triggered by events, which brought short-lived spikes of attention to issues. After the overturning of Roe v Wade, dynamics of information spread among conservatives changed, with issues related to religion, fetal rights and exceptions to abortion ban reverberating among this group. 

\section{Discussion}
We investigated the emotional dimension of political polarization in online networks and the interplay between emotions,  network structure, partisanship  and dynamics of information spread. 
Analyzing emotions expressed in online discussions about abortion and the COVID-19 pandemic, we found that users expressed more emotions in their replies to opposite-ideology users and their valence had the hallmarks of affective polarization, namely ``in-group favoritism, out-group animosity''~\cite{iyengar2012affect}. 

Importantly, we showed that affective polarization generalizes beyond the in-group/out-group dichotomy. When accounting for network distance between interacting users in the retweet network, a proxy of the follower graph, anger, disgust and toxicity increased with distance, while joy largely decreased. These findings generalized across datasets and measures of network distance confirming robustness of findings. 

Our findings shed more light on affective polarization. Only a subset of emotion and toxicity categories we measured displayed group differences: anger, disgust, fear, joy, toxic language and obscenities. Our study also revealed that like joy, fear is usually higher within in-group replies. This stands in contrast to previous studies~\cite{iyengar2015fear}, but highlights the complex role of fear in social interactions. Studies of folklore and mythology suggest that fear helps social cohesion: by making threats salient, fear increases in-group solidarity~\cite{tang2016visualizing,tangherlini2017toward}. 
In-group replies also tend to be longer, supporting the notion that they serve to share information within the group on how to negotiate threats.

One implication of this finding is that information may spread differently among interacting groups within the same population based on its emotional salience to each group. We saw some evidence for this in how liberal and conservative users shared various issues during the COVID-19 pandemic. Conservatives paid more attention to lockdowns during the early phase of the COVID-19 pandemic, as demonstrated by the persistence of  retweets about lockdowns. During this period of time conservatives protested lockdowns, suggesting the emotional importance of this issue in differentiating them from liberals. At other times the lockdowns issue attracted short bursts of attention, triggered by events in the news. Similar patterns in our data point to the complex interplay between emotions,  partisanship and dynamics of information spread within a polarized population.

Our results also highlight important differences between reply and retweet interactions. While researchers sometimes conflate them when building social networks, our findings suggest that these interactions serve a very different purpose and that combining them may obfuscate important features of network structure.

Like any study of social media, ours has limitations that tamper conclusions. By necessity, our datasets represented a small sample of online discussions, even when controlling for the topic. Replication of results in  more representative samples of online discussions could help verify the generalizability of findings. Beyond biases introduced by keyword-centered tweet collection, not observing all interactions may have limited the range of emotions we observed. Similarly, retweets are a biased sample of the follower relationships \cite{rao2023retweets}, which may have impacted our conclusions. Errors in partisanship detection, emotion and toxicity classification, may have further affected our findings. While we cannot discount all of these biases, the consistency of our results across datasets and scenarios gives us confidence about our conclusions.

Our study does not disentangle the directionality of the relationship between network distance and emotive expression, limiting its theoretical contributions. Despite this limitation, we believe that the descriptive analysis of the interplay between emotions and network structure is still valuable.

Another thing to consider is that our results could be explained by some confounder, rather than group polarization or network structure. For example, emotionally charged content is retweeted more frequently \cite{brady2017emotion}, adding emotional texture to retweet networks. Moreover, Twitter's personalization algorithm may highlight  emotionally charged content, thereby driving engagement \cite{brady2021social}, while rapid information spread within communities \cite{weng2013virality} may also distort emotions. Although we cannot discount all confounders, the consistency of our findings across different datasets is encouraging. 

Despite potential limitations like incomplete observability and data bias, the consistency of results across datasets and methods provides confidence in the conclusions of our study about the interplay between emotions, network structure and the dynamics of information spread across and within groups. Understanding these mechanisms is crucial for addressing challenges related to misinformation, polarization, and the health of public discourse in the digital age.

\section{Methods}

\subsection{Data}

We used a public corpus of tweets about the COVID-19 pandemic~\cite{chen2020tracking}, focusing on tweets posted between January 21, 2020 and April 22, 2020 in the analysis of polarization.

Our second dataset is a public corpus of tweets about abortion  rights collected between January 1, 2022 to January 6, 2023~\cite{chang2023roeoverturned}. The tweets contain keywords and hashtags that reflect both sides of the abortion rights debate in US during the period that Roe v Wade was overturned. 

For both datasets, Carmen \cite{dredze2013carmen}, a geo-location tool for Twitter data, was used to link tweets to locations within US. Carmen relies on metadata in tweets, such as ``place'' and ``coordinates'' objects that encode location, as well as  mentions of locations in a user's bio, to infer their location. 
We used this approach to filter out users  whose home location is not one of US states.

To study dynamics of emotional polarization and information spread, we focus on interpersonal interactions in online social networks. On platforms like Twitter, these engagements predominantly manifest through retweets and replies. Each retweet or reply post typically refers back to an original tweet, hereinafter referred to as the parent tweet. Furthermore, every retweet or reply record includes the \verb|ID| of its author as well as the \verb|ID| of the author of the parent tweet. We analyze all retweets and replies where both the author and the referenced tweet's author are situated in the United States and discard the rest. Table \ref{tab:tweet_stats} shows statistics of the resulting datasets.

\subsection{Emotions \& Toxicity Detection}
Emotions represent feelings, which are often  expressed through language. Early attempts to automate emotion recognition from  text relied on emotion lexicons---curated collections of words categorized by their emotional content, e.g., LIWC~\cite{pennebaker2001linguistic}, EmoLex~\cite{mohammad2010emotions}, and WKB~\cite{warriner2013norms}. The advent of transformers has revolutionized emotion detection, which could now benefit from contextual cues. 

To measure emotions  we use an open-source library Demux~\cite{chochlakis2023leveraging}. This model was shown to outperform competing methods on the SemEval 2018 Task 1 e-c benchmark \cite{SemEval2018Task1}. 
Demux assigns none, one or more emotions to input text, along with a scalar value representing its confidence score. The confidence score is the likelihood the tweet expresses that emotion. Demux can recognize a range of emotions in multi-lingual text, including anger, disgust, fear, sadness, joy, love, trust, pessimism and optimism. 

To measure toxicity, we use an open-source classifier Detoxify.\footnote{https://github.com/unitaryai/detoxify} The model is trained on the multilabel toxic comment classification task to recognize toxicity levels of  tweets. The model outputs a score, a scalar value  that captures the likelihood the tweet expresses toxicity, severe toxicity, obscenity, a threat, or an insult. In this study, we only use toxicity scores.

\subsection{Ideology Classification}

To estimate the ideology  of social media users, studies have relied on  follower relationships~\cite{barbera2015birds}, mention and retweet interactions~\cite{conover2011political,badawy2018analyzing}, and URL sharing \cite{le2019measuring, nikolov2020right, cinelli2021echo, rao2021political}. 
Here we use a method described in \cite{rao2021political} to classify individual Twitter users as \textit{liberal} or \textit{conservative} based on the text of the messages they share. The classifier leverages political bias scores assigned to well over 6K online sources by Media Bias-Fact Check (MBFC) \cite{mbfc2023politics}. Based on these scores,  training data is created by assigning each user a score that is a weighted average of the political bias scores of the URLs they share. After training a text embedding-based model on this data, the classifier achieves state-of-the-art performance on recognizing user ideology. 

\subsection{Issue Detection}
Contentious issues that emerged during the pandemic include 1) the \textit{origins of the virus}, involving debates over bats, wet-markets, lab leak and the gain of function research; 2) the implementation of \textit{lockdown} measures via quarantines, business closures,  social distancing and bans on mass congregation; 3) \textit{masking} mandates and face mask shortages; 4) the impact of the pandemic on \textit{education} with school closures and shift to online learning; and 5) \textit{vaccine}-related discussions~\cite{schaeffer2020lab, aidan2021lockdowns, rojas2020masks, pierri2022online, rathje2022social}. 

The issues  central to the abortion debate in the US~\cite{mackenzie2020abortion,pew2022partisan} include 1) \textit{religion} and faith-based arguments against abortion; 2) views promoting primacy of \textit{fetal rights}; 3) framing abortion as a \textit{bodily autonomy} issue and freedom to choose; 4) abortion as a women's \textit{health} issue; and 5) the question of  \textit{exceptions} to abortion restrictions, for example, to save a woman's life or in the case of rape or incest.

We leverage a method developed and validated in previous works to detect these issues. 
 
The method harvests relevant keywords and phrases from Wikipedia pages discussing specific issues, labels a subset of tweets using these terms and trains a transformer-based model on this data. The trained models were shown to achieve state-of-the-art performance recognizing pandemic and abortion-related issues in these datasets~\cite{rao2023pandemic,rao2023tracking}. A tweet could discuss multiple issues or no issue at all.

\subsection{Network Construction}

Studies of online social networks differ in how they represent edges between users. Some researchers~\cite{cinelli2021echo,bollen2011happiness}  use follower relations to capture the attention users pay to others. However, collecting follower links  is highly impractical due to API limitations. Instead, researchers rely on retweets, which can be more easily extracted from the tweets metadata, to construct the social graph~\cite{rao2023retweets}. 
Retweet networks are foundational to social media analysis and have been used in studies of information spread~\cite{Lerman10icwsm}, virality prediction~\cite{weng2013virality}, fake news~\cite{vosoughi2018spread}, online echo chambers~\cite{cinelli2021echo,garimella2018political}, political polarization~\cite{conover2011political,jiang2023retweet}, and online discussions~\cite{wilson2020cross,evkoski2021evolution}.
Following this practice, we construct a retweet-based social network for each dataset. Retweets are evidence that both the author of the original tweet and the author of the retweet were, at least on one occasion,  exposed to the same information. Therefore, we model retweet networks as undirected, unweighted graphs whose nodes represent users and edges represent the existence of at least one retweet between them (in either direction). 

We measure distance in networks in two ways. The first one uses the shortest path between two nodes in the retweet network. The second one measures Euclidean distance between nodes in the embedding space (Figure~\ref{fig:networks}) generated by the LargeVis model. 

\subsection*{Data Availability}
Given the restrictions imposed by X (then Twitter) on publicly sharing tweet objects, the authors state that there are certain restrictions on its availability. Only the tweet identifiers used in this study are publicly available. Readers can apply for API access from X in order to rehydrate these tweets. The tweet identifiers for the COVID-19 dataset are available at \url{https://dataverse.harvard.edu/dataset.xhtml?persistentId=doi:10.7910/DVN/DKOVLA} and ones for the Roe-Wade dataset are available at \url{https://dataverse.harvard.edu/dataset.xhtml?persistentId=doi:10.7910/DVN/STU0J5}.  The DOIs for the COVID-19 and Roe-Wade datasets are: 10.7910/DVN/DKOVLA and 10.7910/DVN/STU0J5 respectively.

\subsection*{Code Availability}
The ideology classifier used to determine individual ideological leanings in this work is available at \url{https://tinyurl.com/yu7xxsey}. The emotion classifier is available at: \url{https://github.com/hasanhuz/SpanEmo}. Toxicity is assessed using the Detoxify model in \url{https://github.com/unitaryai/detoxify}. Geolocation inference is performed using Carmen described in \url{https://github.com/mdredze/carmen}. Code for network visualizations were obtained from \url{https://github.com/lferry007/LargeVis}. Code and auxiliary datasets used as a part of this study are also made available at \url{https://zenodo.org/records/10810851}.

\subsection*{Acknowledgments}

This work was supported in part by AFOSR grant FA9550-20-1-0224, DARPA under contract HR001121C0168 and NSF under grant CCF 2200256.

\subsection*{Author contributions statement}
K.L. conceived the experiment,  A.R. prepared the data, D.F., A.R., K.L. and Z.H. conducted the analysis and analyzed the results.  All authors contributed to writing and reviewing the manuscript.

\subsection*{Competing interests} The authors declare there are no financial or non-financial competing interests.

\bibliography{references}
\clearpage
\section*{Supplementary Information}

\begin{figure*}
    \centering
     {\includegraphics[width=1\linewidth]{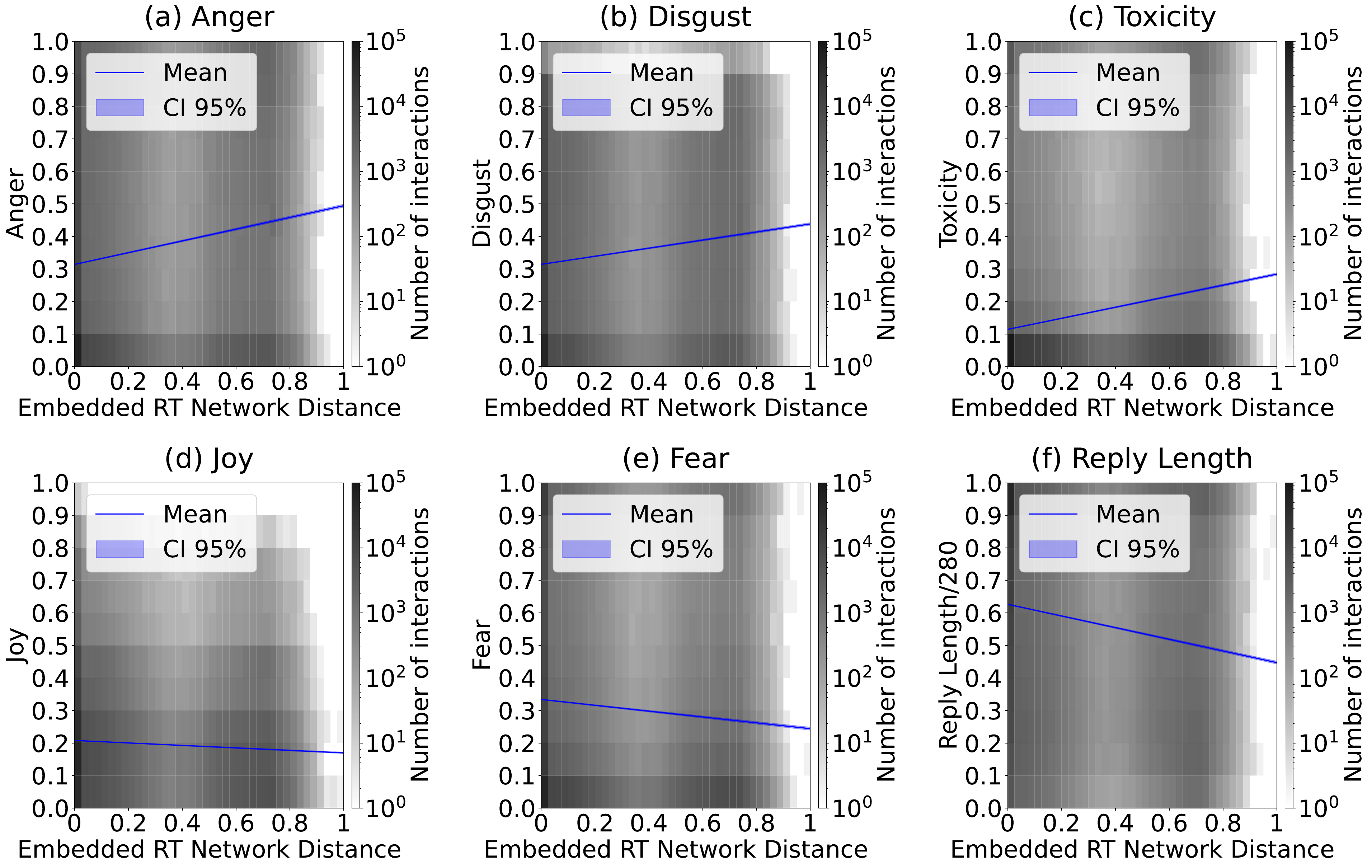}}
    \caption{Affective polarization in the retweet network of Roe\_v\_Wade data. Density plot shows the number of replies  with a specific emotion between two users a given distance apart. Emotions like (a) anger, (b) disgust and  (c) toxicity increase with distance in the network embedding space between users, while (d) joy and (e) fear decrease with distance, as does (f) reply length.}
    \label{fig:distance-rvw}
\end{figure*}

\begin{figure*}
    \centering
     {\includegraphics[width=1\linewidth]{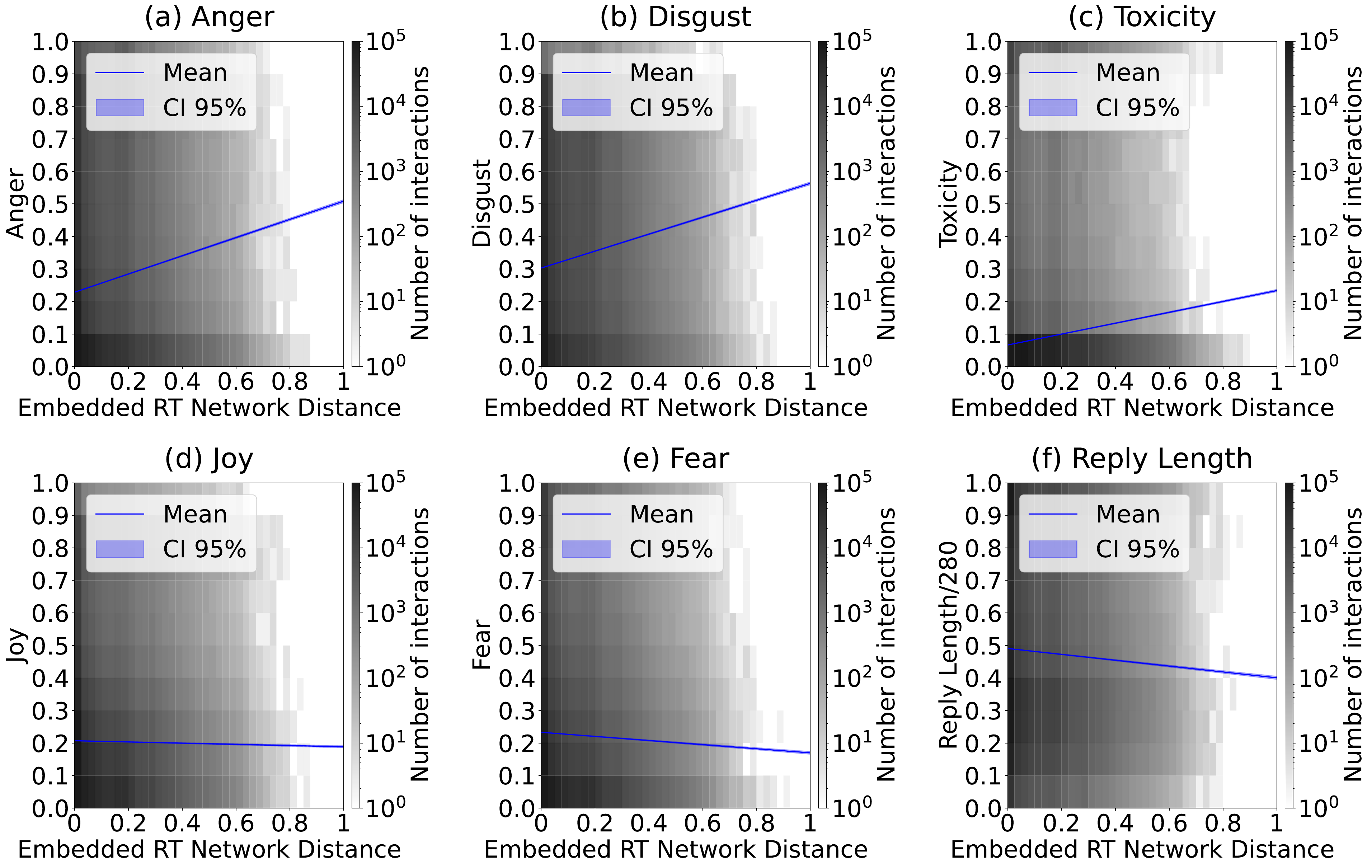}}
    \caption{Affective polarization in the retweet network of COVID-19 data. Density plot shows the number of replies  with a specific emotion between two users a given distance apart. Emotions like (a) anger, (b) disgust and  (c) toxicity increase with distance in the network embedding space between users, while (d) joy and (e) fear decrease with distance, as does (f) reply length.}
    \label{fig:distance-covid}
\end{figure*}

\clearpage

\clearpage

\begin{table*}[hbt!]
\centering
\caption{
            Statistical summary of regression analysis for affect as a function of embedded network distance in COVID-19 RT network between a user replying to another user.  
            }
\label{tab:covid_us_full}
\begin{tabular}{lll}
\toprule
{\textbf{Affect}} & {\textbf{Slope $\pm$ 95\% CI}} & {\textbf{Intercept $\pm$ 95\% CI}} \\
\midrule
Toxicity & 0.1673(0.0032) & 0.0663(0.0005) \\
Anger & 0.28(0.0042) & 0.2286(0.0006) \\
Disgust & 0.2604(0.004) & 0.3029(0.0006) \\
Fear & -0.0632(0.0037) & 0.233(0.0005) \\
Joy & -0.0185(0.003) & 0.207(0.0004) \\
Reply Length/280 & -0.0902(0.0042) & 0.4906(0.0006) \\
No Emotion & -0.1158(0.0028) & 0.2259(0.0004) \\
\bottomrule
\end{tabular}
\end{table*}

\begin{table*}[hbt!]
\centering
\caption{
            Statistical summary of regression analysis for affect as a function of embedded network distance in Roe\_v\_Wade RT network between a user replying to another user.  
            }
\label{tab:roe_wade_us_full}
\begin{tabular}{lll}
\toprule
{\textbf{Affect}} & {\textbf{Slope $\pm$ 95\% CI}} & {\textbf{Intercept $\pm$ 95\% CI}} \\
\midrule
Toxicity & 0.1698(0.0028) & 0.1143(0.0011) \\
Anger & 0.1805(0.0033) & 0.3141(0.0013) \\
Disgust & 0.1243(0.0028) & 0.3142(0.0011) \\
Fear & -0.0899(0.0032) & 0.3339(0.0012) \\
Joy & -0.0382(0.0015) & 0.2079(0.0006) \\
Reply Length/280 & -0.1793(0.0031) & 0.6265(0.0012) \\
No Emotion & -0.0012(0.0016) & 0.1593(0.0006) \\
\bottomrule
\end{tabular}
\end{table*}

\end{document}